\documentclass[12pt]{iopart}

\usepackage{graphicx}
\usepackage{float}
\usepackage{wrapfig}
\usepackage{adjustbox}
\usepackage{subfigure}

\begin{document}



\title{Test particles dynamics in the JOREK 3D non-linear MHD code and application to electron transport in a disruption simulation}

\author{C. Sommariva\textsuperscript{1}, E. Nardon\textsuperscript{1}, P. Beyer\textsuperscript{2}, M. Hoelzl\textsuperscript{3}, G. T. A. Huijsmans\textsuperscript{1,4}, D. van Vugt\textsuperscript{4}, and JET Contributors\textsuperscript{5,*}}

\address{\textsuperscript{1}CEA, IRFM, F-13108, Saint Paul-lez-Durance, France}
\address{\textsuperscript{2}Aix-Marseille Universit\'e, CNRS, PIIM UMR 7345, 13397, Marseille Cedex 20, France}
\address{\textsuperscript{3}Max-Planck Institute for Plasma Physics, Boltzmannstr. 2, 85768, Garching, Germany}
\address{\textsuperscript{4}Dep. Applied Physics, T. U. Eindhoven, P. O. B. 513, 5600, Eindhoven, Netherlands}
\address{\textsuperscript{5}EUROfusion Consortium, JET, Culham Science Center, Abingdon, OX14 3DB, UK}
\address{\textsuperscript{*}See the author of 'Overview of the JET results in support to ITER', by X. Litaudon et al., to be published in Nuclear Fusion special issue: overview and summary reports from the 26\textsuperscript{th} Fusion Energy Conference (Kyoto, Japan, 17-22 October 2016)}
\ead{cristian.sommariva@cea.fr}
\vspace{10pt}
\begin{indented}
\item[]September 2016
\vspace{2pc}
\submitto{\NF}
\end{indented}
\maketitle

\begin{abstract}
In order to contribute to the understanding of runaway electron generation mechanisms during tokamak disruptions, a test particle tracker is introduced in the JOREK 3D non-linear MHD code, able to compute both full and guiding center relativistic orbits. Tests of the module show good conservation of the invariants of motion and consistency between full orbit and guiding center solutions. A first application is presented where test electron confinement properties are investigated in a massive gas injection-triggered disruption simulation in JET-like geometry. It is found that electron populations initialised before the thermal quench (TQ) are typically not fully deconfined in spite of the global stochasticity of the magnetic field during the TQ. The fraction of ``survivors'' decreases from a few tens down to a few tenths of percent as the electron energy varies from 1keV to 10MeV. The underlying mechanism for electron ``survival'' is the prompt reformation of closed magnetic surfaces at the plasma core and, to a smaller extent, the subsequent reappearance of a magnetic surface at the edge. It is also found that electrons are less deconfined at 10MeV than at 1MeV, which appears consistent with a phase averaging effect due to orbit shifts at high energy.
\end{abstract}


\section{Introduction} \label{introduction}


Runaway Electrons (RE) are defined as the fraction of a plasma electron population being in the phase space region where the drag force due to collisions does not compensate the acceleration caused by a driving electric field \cite{dreicer59}. RE beams carrying a substantial fraction of the plasma current can be generated during a tokamak disruption. When such an event happens in large machines, for example in JET \cite{reux15}, the plasma facing components struck by the terminating RE beam suffer high heat loads, sometimes leading to melting or sputtering \cite{reux15}. ITER is designed to generate plasmas having high currents ($\mathrm{I_p \approx 15MA}$) which implies that a larger energy may be contained in RE beams. RE impacts are therefore a threat to the successful operation of ITER and prevention or mitigation techniques are likely to be necessary \cite{sugihara12}\cite{lehnen14}. The design of such techniques should ideally rely on a good understanding of the mechanisms underlying the generation and dissipation of RE.

A typical disruption \cite{hender07} comprises two main consecutive phases: the Thermal Quench (TQ), during which the thermal energy of the plasma is lost over a millisecond timescale, and the Current Quench (CQ), i.e. the fast decay of the plasma current due to the very large post-TQ plasma resistance, which terminates the discharge. During the CQ, a strong self-induced toroidal electric field appears in the plasma which is typically large enough to give rise to a runaway avalanche, i.e. an exponentiation of the number of RE due to knock-on collisions of RE onto thermal electrons \cite{rosenbluth97}. The avalanche mechanism however needs an initial ``seed'' RE population and the origin of this seed is less clear. According to \cite{martin-solis17} (for ITER) and \cite{reuxPhD} (for Tore Supra), a direct acceleration of thermal electrons by the electric field, the so-called ``Dreicer mechanism'' \cite{dreicer59}\cite{connor75}, is unlikely to take place during the CQ because the electric field is not large enough. A more likely candidate is the so-called ``hot tail'' mechanism \cite{smith05}\cite{smith08}, which relies on the fact that electrons from the high energy tail of the pre-TQ distribution take a longer time to thermalise than the TQ duration. As a consequence, these electrons are still ``hot'' at the beginning of the CQ and may therefore be accelerated by the electric field and become RE. Another possibility is that RE may be formed by the Dreicer mechanism already during the TQ \cite{helander04}\cite{plyusnin06}\cite{boozer17}.
 
The last two above-mentioned mechanisms can however be envisaged only if the relevant fast electrons remain (at least partly) confined throughout the TQ. This is questionable because it is thought that magnetohydrodynamic (MHD) activity during the TQ makes the magnetic field stochastic over a significant part (if not all) of the plasma volume. Due to their fast motion along field lines, these electrons may therefore be expected to be lost. However, MHD fluctuations decay after the TQ so that flux surfaces may promptly reappear (as indeed observed in simulations, see below, and also as suggested experimentally by means of a tomographic reconstruction of soft X-ray data \cite{plyusnin06}) and stop the loss process. It is therefore difficult to conclude \textit{a priori} on how much of the fast electrons eventually remain in the plasma. This is the main question addressed in the present paper. It should be noted that, due to the extremely large avalanche amplification factor expected in ITER (present theoretical estimations suggest that up to 40 e-folds may be possible \cite{boozer17}) and to the fact that only $\sim 10^{19}$ RE would be sufficient to carry $10$MA in ITER, even tiny fractions of ``survivors'' many orders of magnitude below 1\% could make a very significant difference. 

The consequences of magnetic stochasticity on fast electron transport have been explored in several theoretical \cite{abdullaev14}\cite{boozer15}\cite{boozer16}\cite{boozer162}\cite{abdullaev16}, numerical \cite{papp111}\cite{papp112}\cite{sarkimaki16}, and experimental \cite{abdullaev16} works, but not necessarily in disruptive situations. From the experimental point of view and in a disruptive context, a clear trend has been observed in some machines toward smaller RE currents as MHD fluctuations during (and just after) the TQ get stronger \cite{zeng13}\cite{zeng17}. Regarding numerical modelling, to our knowledge the only studies on fast electron transport in fields from disruptions simulated by a 3D non-linear MHD code have been performed by Izzo et al. with the NIMROD code \cite{izzo11}\cite{izzo12}.

In view of addressing the above questions, we introduce in the 3D non-linear MHD code JOREK \cite{huysmans07}\cite{czarny08} a module capable of computing relativistic test electron full and guiding center orbits. In the present paper, attention is given to the description of the new module (Section \ref{sec_code_overview}) and its tests (Section \ref{numerical_tests}). A first physical application to test electron confinement in a JOREK-simulated JET disruption triggered by a Massive Gas Injection (MGI) is then presented (Section \ref{physical_results}). Section \ref{conclusions} discusses the results and future plans.
\section{A relativistic test particle module in JOREK} \label{sec_code_overview}

\subsection{Full orbit model} \label{fullorbit}

The dynamics of relativistic particles is described by the following set of equations, called the Full Orbit (FO) model \cite{landau71}:

\numparts
\begin{eqnarray}
\dot{\mathbf{x}} = \frac{\mathbf{p}}{m\gamma} \label{particle_position} \\
\dot{\mathbf{p}} = q\left({\mathbf{E}} + {\frac{\mathbf{p}}{m\gamma}}{\times}{\mathbf{B}}\right) \label{particle_momentum} \\
\gamma = \sqrt{1+\left(\frac{\mathbf{p}}{mc}\right)^2} \label{gamma_particle}, \quad \mathbf{p}^2=\mathbf{p} \cdot \mathbf{p}
\end{eqnarray}
\endnumparts

where $q$ is the particle charge, $m$ its rest mass, $c$ the speed of light and $\mathbf{E}$ and $\mathbf{B}$ the electric and magnetic fields. A numerical computation of the particle trajectory requires the resolution of the gyromotion \cite{landau71}\cite{hazeltine03} which implies a time step of the order of $\mathrm{{\Delta}t=0.01 \cdot T_{gyro}}$ where $\mathrm{T_{gyro}={2\pi}\frac{m \gamma}{|{q}|B}}$ is the gyration period (in this work $\mathrm{T_{gyro}=T_{gyro}|_{\gamma=1}={2\pi}\frac{m}{|{q}|B}}$ will be implicitly used) \cite{landau71}, i.e. ${\Delta}t \sim10^{-13}s$ for an electron in a 2T magnetic field. This means that a few $10^{10}$ iterations are needed in order to track an electron for a few milliseconds, which may result in a poor solution accuracy if a non-conservative scheme is used \cite{qin08}. Therefore the Volume Preserving Algorithm (VPA), a symplectic scheme developed in \cite{zhang15}\cite{wang16}\cite{he16}, is implemented. For completeness, the VPA scheme for the $\mathrm{k_{th}}$ iteration is reported hereafter:

\numparts
\begin{eqnarray}
{\mathbf{x}}_{k+\frac{1}{2}} = {\mathbf{x}}_{k} +{\frac{\Delta t}{2m}}{\frac{\mathbf{p}_{k}}{\sqrt{1+\left(\frac{\mathbf{p}_k}{mc}\right)^2}}} \\
\hat{\mathbf{p}} = {\mathbf{p}_{k}} + {\frac{q\Delta t}{2}}{\mathbf{E}_{k+\frac{1}{2}}} \\
\overline{\mathbf{p}} = {\mathrm{Cay}}\left({\frac{q\Delta t}{2m}}{\frac{{_{\times}}\underline{\underline{B}}_{k+\frac{1}{2}}}{\sqrt{1+(\frac{\hat{\mathbf{p}}}{mc})^2}}}\right)\cdot{\hat{\mathbf{p}}} \label{particle_B_momentum} \\
{\mathbf{p}}_{k+1} = \overline{\mathbf{p}} + {\frac{q\Delta t}{2}}{\mathbf{E}_{k+\frac{1}{2}}} \\
{\mathbf{x}}_{k+1} = {\mathbf{x}}_{k+\frac{1}{2}} +{\frac{\Delta t}{2m}}{\frac{\mathbf{p}_{k+1}}{\sqrt{1+\left(\frac{\mathbf{p}_{k+1}}{mc}\right)^2}}} \\
{_{\times}}\underline{\underline{B}} = \left [ \begin{array}{c c c}
\mathrm{0} & \mathrm{B_{3}} & \mathrm{-B_{2}} \\
\mathrm{-B_{3}} & \mathrm{0} & \mathrm{B_{1}} \\
\mathrm{B_{2}} & \mathrm{-B_{1}} & \mathrm{0}
\end{array} \right ]
\end{eqnarray}
\endnumparts

where $\mathrm{B_{j}}$ is the $\mathrm{j_{th}}$ component of the magnetic field vector, ${\mathrm{Cay}}\left(\underline{\underline{A}}\right)={\left(\underline{\underline{I}}-\underline{\underline{A}}\right)^{-1}}\left(I+\underline{\underline{A}}\right)$ is the Cayley transform of the matrix $\underline{\underline{A}}$, and $\underline{\underline{I}}$ is the identity matrix. It is important to stress that the equations of motion are resolved in a Cartesian reference system in order to preserve symplecticness. The perfect conservation of the symplectic structure implies that deviations of the invariants of motions are bounded, in accordance with the scheme order and time step, and do not drift (for a deeper insight into this subject see \cite{morrison17} and references therein). In reality, numerical errors arising from the description of the plasma fields (such as the finite numerical solution smoothness and accuracy) and from the particle tracking procedure (described in Section \ref{tracking_procedure}) inevitably break the conservation of the symplectic structure. Thus, a perfect bounding of the invariants of motion should not be expected.

Finally, we remark that in the following, we use the exact Cayley transform instead of the computationally less expensive approximated form proposed in \cite{zhang15}. This choice is based on a comparison between the two methods, computing the orbit of an electron with a kinetic energy of 10MeV and a pitch angle of $170^{\circ}$ in an equilibrium plasma field. The total simulation time and time step were respectively of $\mathrm{T=1ms}$ and $\mathrm{\Delta t=1.4 \cdot 10^{-2} \cdot T_{gyro}}$. Despite a $27\%$ reduction of computational time (average on 10 tests), the approximated Cayley transform was found to perform much worse regarding the conservation of the canonical toroidal momentum ($\mathrm{P_{\phi}}$) with fluctuations having an amplitude up to $2.5\%$, while the exact Cayley transform showed a $\mathrm{P_{\phi}}$ conservation error of only $2.8 \cdot 10^{-7} \%$ (the total energy conservation errors of the two methods were similar).

\subsection{Guiding center model} \label{guidingCenter}

The Guiding Center (GC) model used in the JOREK particle tracker is the first order energy-like relativistic GC model described in \cite{tao07} and \cite{cary09}:

\numparts
\begin{eqnarray}
\eqalign{\mathbf{\dot{X}} = {\frac{1}{{\mathbf{b}} \cdot {\mathbf{B^{\ast}}}}} \left({q{\mathbf{E}}{\times}{\mathbf{b}}} - {{p_{\parallel}}{\frac{{\partial}{\mathbf{b}}}{\partial t}} \times {\mathbf{b}}} + {\frac{m{\mu}{\mathbf{b}}{\times}{\mathbf{{\nabla}B}} + {p_{\parallel}}{\mathbf{B^{\ast}}}}{m \gamma_{GC}}}\right)} \label{gc_position}  \\
\eqalign{\dot{p_{\parallel}} = {\frac{\mathbf{B^{\ast}}}{\mathbf{b} \cdot \mathbf{B^{\ast}}}} \cdot \left(q{\mathbf{E}}-{{p_{\parallel}}{\frac{{\partial}{\mathbf{b}}}{\partial t}}}-{\frac{{\mu}{\nabla}{B}}{\gamma_{GC}}}\right)} \label{gc_momentum} \\
\gamma_{GC} = \sqrt{1+{\left(\frac{p_{\parallel}}{mc}\right)^2} + {\frac{2{\mu}B}{mc^2}}} \label{gamma_GC}
\end{eqnarray}
\endnumparts

where ${\mathbf{X}}$ is the GC position vector, $p_{\parallel}$ is the particle momentum in the direction of the magnetic field, ${\mu}=\frac{{\| \mathbf{p}-{p_{\parallel}}\mathbf{b} \|}^2}{2mB}$ is the magnetic moment \cite{tao07}, $B$ is the magnetic field norm, ${\mathbf{b}}={\frac{\mathbf{B}}{B}}$ the magnetic field direction and ${\mathbf{B^{\ast}}}={p_{\parallel}}{{\mathbf{\nabla}}\times {\mathbf{b}}}+q{\mathbf{B}}$ is the so-called ``effective magnetic field''. Being a reduced dynamical model, the first order GC approximation is associated to a number of validity conditions:
\begin{enumerate}
\item[1] The electromagnetic field time scale T has to be much longer than the particle gyro-period $T_{gyro}$: $\frac{T_{gyro}}{T} \ll 1$, with $T_{gyro}$ defined in subsection \ref{fullorbit}.
\item[2] The electromagnetic field length scale L has to be much larger than the gyro-radius $\rho$: $\frac{\rho}{L} \ll 1$, where ${\rho}=\frac{{\|}\vec{p}-{p_{\parallel}}{\hat{b}}{\|}}{|q|B}$ \cite{landau71}.
\item[3] The particle displacement in a gyro-period along $\mathbf{b}$ has to be small compared to the electromagnetic field parallel variation length scale $L_{\parallel}$: $\frac{l_{\parallel}}{L_{\parallel}} \ll 1$, where $l_{\parallel}=\frac{2{\pi}|p_{\parallel}|}{|q|B}$ is an estimate from equation \ref{gc_position} of the GC parallel displacement.
\item[4] The electric field has to satisfy $\frac{|{E_{\parallel}}|}{{{E}_{\perp}}}\sim{\frac{\rho}{L}} \ll 1$, having defined ${E_{\parallel}}={\mathbf{E}}{\cdot}{\mathbf{b}}$ and ${E}_{\perp}={\| \mathbf{E}-{E_{\parallel}}{\mathbf{b}} \|}$. This condition is related to the GC ordering self-consistency as discussed in \cite{littlejohn81}\cite{cary09}.
\end{enumerate}

In order to make estimates regarding conditions 1, 2 and 3 in a RE context, the length and time scales for electrons with energies between 0.5 and 500MeV, as well as the order of magnitude of the ratios $\frac{T_{gyro}}{T}$, $\frac{\rho}{L}$ and $\frac{l_{\parallel}}{L_{\parallel}}$ are given in Table \ref{tab_part_chart}. These values assume a magnetic field of 2T and a tokamak major radius of 3m and minor radius of 1m (i.e. JET-like parameters). Here, $L$ and $L_{\parallel}$ are calculated as the minimum gradient lengths ($L=\min\frac{B}{\|{\mathbf{\nabla B}}\|}$ and $L_{\parallel}=\min\frac{B}{|\mathbf{b}\cdot \mathbf{\nabla B}|}$) of a simple axisymmetric equilibrium tokamak-like magnetic field with a constant q=1 profile (estimations in fields from JOREK disruption simulations are provided in Section \ref{numerical_tests}). Their values are $L\mathrm{=2.66m}$ and $L_{\parallel}\mathrm{=5.22m}$. The plasma characteristic time T is conservatively taken to be the smallest JOREK time step used in the most extreme disruption simulations: T$=3\cdot 10^{-8}$s (a typical JOREK time step is within $\mathrm{\sim [10^{-7},10^{-5}]}$s). The Larmor radius is obtained using the limiting pitch angle of 90 degrees while 0 degrees is used for the parallel displacement. Table \ref{tab_part_chart} indicates that the GC model should be a good approximation for electrons up to a few MeV while a numerical assessment is required for energies of tens of MeV. For energies above hundreds of MeV, FO simulations are advisable \cite{wang16}. Numerical assessments are also required concerning the validity of condition 4 since it is very hard to estimate the electric field \textit{a priori}. 

\begin{table}[h]
\centering
\begin{tabular}{| c | c | c | c | c | c | c |}
\hline
$\mathrm{E_{kin}}$ (MeV) & $\mathrm{T_{gyro}}$ (ns) & $\mathrm{\rho}$ (mm) & $\mathrm{l_{\parallel}}$ (mm) & $\mathrm{OfM \left( T_{gyro}/T \right)}$ & $\mathrm{OfM \left( \rho/L \right)}$ & $\mathrm{OfM \left(l_{\parallel}/L_{\parallel} \right)}$ \\
\hline
0.5 & 0.035 & 1.45 & 9.14  & $10^{-4}$ & $10^{-4}$ & $10^{-3}$ \\
\hline
5 & 0.19 & 9.15 & 57.5 & $10^{-3}$ & $10^{-3}$ & $10^{-2}$ \\
\hline
50 & 1.77 & 84.2 & 529 & $10^{-2}$ & $10^{-2}$ & $10^{-1}$ \\
\hline
500 & 17.5 & 835 & 5245 & $10^{-1}$ & $10^{-1}$ & $10^0$ \\
\hline
\end{tabular} 
\caption{Electron length and time scale estimates, as well as the order of magnitude (denoted OfM) of the critical ratios for GC validity, at different energies (see text for details)}
\label{tab_part_chart}
\end{table}

In terms of numerical scheme, the GC equations are solved using the fifth order Cash-Karp Runge-Kutta scheme described in \cite{cash90}, using a cylindrical $(R,Z,\phi)$ coordinate system, where $\phi$ is the toroidal angle. An adaptive time stepping (allowing only for a time step \textit{reduction} from the one set by the user) has been implemented in order to mitigate, if need be, the lack of symplecticness \cite{qin08}. This is based on the truncation error control method, originally developed in \cite{shampine77}\cite{shampine79} and reported in \cite{press97}, using the total energy and the canonical toroidal momentum as control variables. These variables were chosen because they are conserved quantities in stationary axisymmetric fields. In what follows, the $\Delta t$ values are however chosen quite small so that the controller action is almost negligible.

\subsection{JOREK fields description} \label{field_interp}

JOREK uses a finite element method to compute the plasma evolution in realistic tokamak geometries \cite{huysmans07}. In the reduced MHD version of JOREK, the field variations in the poloidal plane are described using a B{\'e}zier surface for each quadrangular mesh element \cite{czarny08}. In the toroidal direction, a Fourier expansion is used. A generic interpolant of a JOREK variable (here noted $\psi$) has the following representation:
\numparts
\begin{eqnarray}
\psi{\left(t,r,s,\phi\right)} = \displaystyle \sum_{k=0}^{N} \left[{\Psi_{k}}{\left(t,r,s\right)}{\cos}\left(k\phi\right)+{{\bar{\Psi}}_{k}{\left(t,r,s\right)}}{\sin}\left(k\phi\right)\right] \\
A_{k}{\left(t,r,s\right)} = \sum_{i,j=0}^{3} {\hat{A}}_{k,i,j}{\left(t\right)}{B_{i}^{3}}\left(r\right){B_{j}^{3}\left(s\right)} \textnormal{ with } A_{k} = \Psi_{k} \textnormal{ or } \bar{\Psi}_{k} \label{eq:4b}  \\
{B_{n}^{3}}\left(p\right) = {\frac{3!}{n!\left(3-n\right)!}}{p^n}{\left(1-p\right)^{3-n}} 
\end{eqnarray}
\endnumparts

where $t$ is the time and $\{r,s\}\in[0,1]\times[0,1]$ are the mesh element local coordinates.  An important feature of this representation is the $\mathit{C^{1}}$ continuity which comes from coefficient constraints typical of B{\'e}zier surface representations \cite{czarny08}. A new feature introduced within the JOREK particle module is the time interpolation of fields, which is necessary due to the particle time step being typically orders of magnitude smaller than the JOREK one. The Hermite-Birkhoff interpolant, which is used for all the simulations presented in this paper, is given, for $t_0 \leq t \leq t_1$ (where $t_0$ and $t_1$ denote two successive JOREK times), by \cite{quarteroni07}:

\numparts
\begin{eqnarray*}
{\tilde{A}}_{k,i,j}{\left(t\right)} = \displaystyle \sum_{m=0}^{1} \left[{\hat{A}}_{k,i,j}{\left(t_m\right)}C_m \left( t \right)+\frac{\mathrm{d} {{\hat{A}}_{k,i,j}}}{\mathrm{d}t} \left(t_m \right) D_m \left( t \right)\right] \\ \quad \quad \quad \quad \quad \textnormal{(where ${\hat{A}}_{k,i,j}$ are the same as in Eq. \ref{eq:4b})}\\
C_m \left( t \right) = [1-2\left(t-t_m\right){\frac{\mathrm{d} l_m}{\mathrm{d}t}}\left(t_m\right)]{l_m^2}\left(t\right) \\
D_m \left( t \right) = \left(t-t_m\right){l_m^2}\left(t\right) \\
l_m \left( t \right) = \displaystyle \sum_{o=0,o{\neq}m}^{1} {\frac{t - t_o}{t_m - t_o}}
\end{eqnarray*}
\endnumparts

The Hermite-Birkhoff interpolant is chosen for its properties of locality (the interpolation in the time interval $[t_0,t_1]$ depends only on the JOREK solution at $t_0$ and $t_1$) and continuity (the global interpolation has $\mathit{C^{1}}$ continuity). Altogether, the use of B{\'e}zier (Bernstein) and Fourier polynomials and Hermite-Birkhoff interpolants guarantees global $\mathit{C^{1}}$ continuity in space and time of both electric and vector potentials, which translates to $\mathit{C^{0}}$ continuity for the electric and magnetic fields.





\subsection{Particle tracking in the JOREK mesh} \label{tracking_procedure}
 
As mentioned above, in the FO (resp. GC) model, particles are pushed in a global Cartesian (resp. cylindrical) coordinate system. On the other hand, plasma fields are given in mesh element local coordinates (see Section \ref{field_interp}). Therefore a procedure to find the element and the local coordinates $\{r,s\}$ corresponding to the particle $\{R,Z\}$ position is required (after a conversion from Cartesian to cylindrical coordinates in the FO case). This identification, called ``particle tracking'', is achieved in the following way. 

First, the new particle position is sought in the local coordinate system of the element where the particle was at the previous time step. Due to the fact that the inverse of a B{\'e}zier interpolant is not known in closed form, a Newton algorithm with backtracking is used. Denoting $k$ the Newton iteration index, we define $\Delta{\left\{R,Z\right\}_{k}}=\left\{R,Z\right\}-{\left\{R,Z\right\}_{k}}$ where $\left\{R,Z\right\}_{k}$ is obtained from  $\left\{r,s\right\}_{k}$ by B{\'e}zier interpolation. A first estimation of the increment of the particle local coordinates at the ${\mathrm{k_{th}}}$ Newton iteration is calculated as:

\begin{equation}
\left[
\begin{matrix}
{\Delta r_k},{\Delta s_k}
\end{matrix}
\right]^T
=
\left[{J(r,s)_{k-1}}\right]^{-1}
\left[
\begin{matrix}
{\Delta R_{k-1}},{\Delta Z_{k-1}}
\end{matrix}
\right]^T
\end{equation}

where $\left[{J(r,s)_{k-1}}\right]^{-1}$ is the inverse of the Jacobian matrix of the $\left\{r,s\right\}\to\left\{R,Z\right\}$ coordinate transformation obtained at the ${\mathrm{\left(k-1\right)_{th}}}$ Newton iteration. In order to increase the convergence rate, a backtracking method is used. Denoting $m$ the backtracking loop index, an estimate of the particle position in local coordinates is calculated as:

 \begin{equation}
\left\{r,s\right\}_{k,m}=\left\{r,s\right\}_{k-1} + {0.5^{m-1}}\left\{\Delta r,\Delta s\right\}_{k}
\end{equation}

The backtracking loop terminates when the error $\mathrm{err}_{k,m}$ at the present backtracking iteration is smaller than the one at the ${\mathrm{\left(k-1\right)_{th}}}$ Newton iteration $\mathrm{err}_{k-1}$ , having defined: $\mathrm{err}_{k,m} = \sqrt{\left(R-R_{k,m}\right)^2 + \left(Z-Z_{k,m}\right)^2}$ and $\mathrm{err}_{k-1} = \sqrt{\left(R-R_{k-1}\right)^2 + \left(Z-Z_{k-1}\right)^2}$. The Newton algorithm stops when the error between two Newton iterations is below a pre-set tolerance or when a pre-set  maximum number of iterations is reached.

If the converged $\left\{r,s\right\}_{k}$ is not inside $\left[0,1\right]\times \left[0,1\right]$, which means that the particle has changed element, a logic verifies which side of the element has been crossed and its associated neighbouring element (which is known thanks to a mesh pre-processing)   is selected. The Newton method with backtracking is then repeated in the new element using a position guess obtained from a first order estimation of the old element origin in new element coordinates (also computed in the pre-processing). If convergence is not reached then the particle is searched in each mesh element using a standard Newton method.

\subsection{Test particles initialisation} \label{jorek_fast_part_init}

Test particles are typically initialised via a Monte Carlo method where the sequences of random numbers are obtained using the {\it{PCG}} random number generator \cite{oneillYY}. In physical space, particles are sampled from uniform distributions in Z and $\mathrm{\phi}$ coordinates while a correction is used for the radial position in order to guarantee uniform particle density ($\mathrm{R = \sqrt{{{R_{min}}^2} + {N_s}\left({{R_{max}}^2}-{{R_{min}}^2}\right)}}$ where $\mathrm{N_s \in \left[0,1\right]}$ is a random number from a constant uniform distribution). A standard acceptance-rejection procedure \cite{birdsall91} may then be performed in order to initialise particle populations in a restricted region of space, for example near a given flux surface, as done in Section \ref{physical_results}.

Velocity space initialisation is obtained via Monte Carlo sampling from uniform distributions of kinetic energy, pitch angle $\theta$ $\mathrm{\left({\cos{\theta}} = \frac{\vec{p} \cdot \hat{b}}{\|{\vec{p}}\|}\right)}$ and (for FO only) gyroangle ($\mathrm{{\tan{\chi}} = \frac{\vec{p} \cdot {\hat{e}}_{\perp}}{\vec{p} \cdot {\hat{e}}_{\nabla \psi}}}$, having defined $\mathrm{{\hat{e}}_{\nabla \psi} = \frac{\nabla \psi - \left(\nabla \psi \cdot \hat{b}\right)}{\| \nabla \psi - \left(\nabla \psi \cdot \hat{b}\right) \|}}$ and $\mathrm{{\hat{e}}_{\perp} = \hat{b} \times {\hat{e}}_{\nabla \psi}}$). Moreover, the code has the possibility to initialise GC from particles, allowing comparisons between models. This is done by computing the GC position using the first order GC transformation and its velocity by matching particle total energy and toroidal canonical momentum as described in \cite{pfefferle15}:

\numparts
\begin{eqnarray}
\vec{x}^{GC} = \vec{x} - \frac{\vec{B}}{qB^2} \times \vec{p} \\
p^{GC}_{\parallel} = \frac{1}{p_{\phi}b_{\phi}}\left[Rp_{\phi}+q\left(\psi - \psi^{GC}\right)\right] \\
\mu^{GC} = \frac{E_{0}}{2B^{GC}}\left\{ \left[\gamma + \frac{q}{E_0} \left(\Phi - \Phi^{GC} \right)\right]^2 - \left[1+\left( \frac{p^{GC}_{\parallel}}{mc}\right)^2 \right] \right\}
\end{eqnarray}
\endnumparts

where $\left( \ast \right)^{GC}$ are GC quantities, $\left( \ast \right)_{\phi}$ vector components along the toroidal direction, $E_0=mc^2$ is the particle rest energy and $\Phi$ is the electric potential. It is necessary to point out that particle and GC magnetic moments are not strictly equal (enforcing such equality results in a overconstrained problem). This implies that their orbits will present a (generally) small mismatch.

\section{Numerical tests} \label{numerical_tests}
In this section, tests of the JOREK test particle module for both FO and GC models are presented.
Orbits are computed in fields from a JOREK disruption simulation of JET pulse 86887, an ohmic pulse with $\mathrm{B_t=2 T}$, $\mathrm{I_p=2 MA}$, $\mathrm{q_{95}=2.9}$ where a disruption was triggered using a $\mathrm{D_2}$ MGI \cite{nardon17}\cite{fil2015}. Electrons are followed for a physical time of 1ms in equilibrium (axisymmetric) or pre-disruptive (non-axisymmetric) MHD fields which are kept stationary in time. This allows testing the conservation of, in the first case, kinetic energy $\mathrm{E_{kin}}$ and toroidal canonical momentum $\mathrm{P_{\phi}}$, and in the second case, total energy $\mathrm{E_{tot}}$. In each set of fields, both a passing relativistic electron ($\mathrm{E_{kin}=10MeV}$, $\mathrm{\theta = 170^{\circ}}$, $\mathrm{\chi = 0^{\circ}}$) and a trapped relativistic electron ($\mathrm{E_{kin}=1MeV}$, $\mathrm{\theta = 100^{\circ}}$, $\mathrm{\chi = 0^{\circ}}$) are tracked. A scan in time step $\Delta t$ is performed to verify the solution convergence. A numerical assessment of the GC validity conditions (see Section \ref{guidingCenter}) is also performed.

\subsection{Tests in stationary axisymmetric fields} \label{eq_verification}

\begin{figure}[h]
\centering
	\subfigure[A][Global view] {\includegraphics[width=5.0cm, height=8cm]{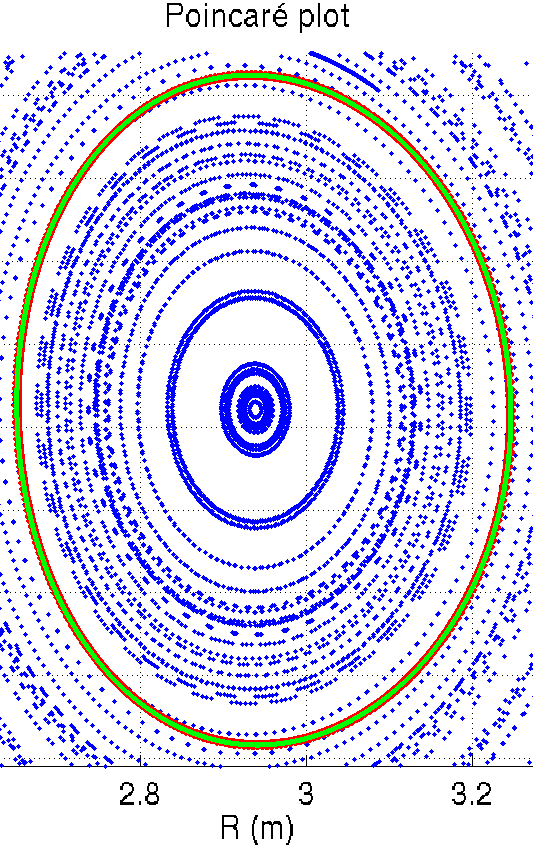} }
	\subfigure[B][Zoom] {\includegraphics[width=2.8572cm, height=8cm]{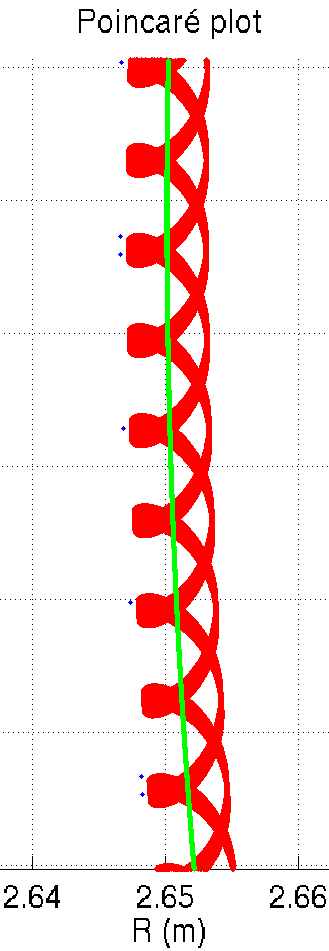} }
\caption{Passing orbits in stationary axisymmetric fields: red dots, green dots and blue dots are respectively FO, GC and field line solutions}
\label{fig:eq_core_passing_part}
\end{figure}

\begin{figure}[h]
\centering
	\subfigure[A][Global view] {\includegraphics[width=2.42853cm, height=8cm]{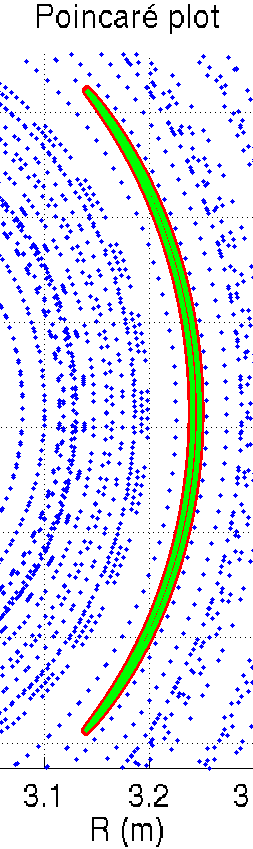} }
	\subfigure[B][Zoom on banana tip] {\includegraphics[width=8cm, height=8cm]{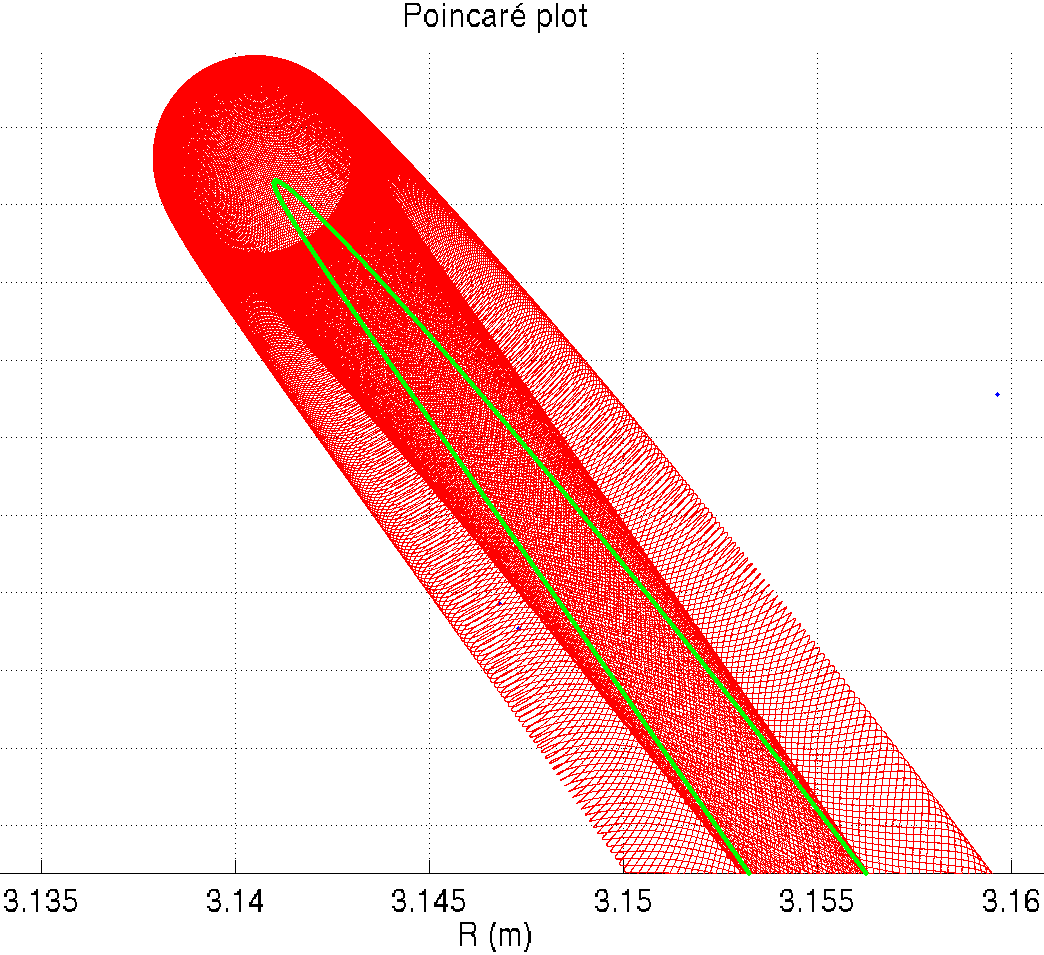} }
\caption{Trapped particle in stationary axisymmetric fields: red dots, green dots and blue dots are respectively FO, GC and field line solutions}
\label{fig:eq_core_trapped_part}
\end{figure}

Figures \ref{fig:eq_core_passing_part} and \ref{fig:eq_core_trapped_part} represent the solutions for respectively passing and trapped orbits in the core region (initial position: R=3.25m, Z=0.22m, $\mathrm{\phi=45^{\circ}}$) for stationary axisymmetric fields.

It can be seen in Figure \ref{fig:eq_core_passing_part} that the passing electron describes a toroidal surface (red and green dots for FO and GC, respectively), called drift surface, which is shifted radially outward compared to magnetic surfaces (blue dots). This shift is related to the grad-B and curvature drifts which play an important role at high energy \cite{goldston95}\cite{abdullaev14}\cite{brizard11}. Figure \ref{fig:eq_core_trapped_part} shows the banana orbit typical of trapped particles \cite{abdullaev14}\cite{brizard11}. Zooms on small parts of the orbits (right plots in Figures \ref{fig:eq_core_passing_part} and \ref{fig:eq_core_trapped_part}) show the precise consistency between the FO and GC trajectories.

Conservation properties are assessed both for the above test cases, where electrons are located in the core region, as well as for cases with electrons near the edge (initial position: R=2.98m, Z=1.3m, $\mathrm{\phi=45deg}$). This is important because in this JOREK simulation, the mesh is coarser at the edge than in the core. 

With the FO tracker, the kinetic energy $\mathrm{E_{kin}}$ is conserved within $10^{-9} \%$ in all test cases, independently of $\mathrm{\Delta t}$ (as long as $\mathrm{\Delta t \leq 1.4 \cdot 10^{-1} \cdot T_{gyro}}$ - note that above this time step, the gyromotion is not well reproduced). For $\mathrm{\Delta t = 1.4 \cdot 10^{-3} \cdot T_{gyro}}$, the canonical toroidal momentum $\mathrm{P_{\phi}}$ is conserved within $10^{-8} \%$ in the core and within $10^{-6} \%$ in the edge. The difference may be related to the coarser mesh at the edge. In contrast to $\mathrm{E_{kin}}$, the $\mathrm{P_{\phi}}$ conservation degrades with increasing $\mathrm{\Delta t}$, roughly quadratically.

\begin{figure}[h]
\centering
\includegraphics[width=12.5cm, height=6.5cm]{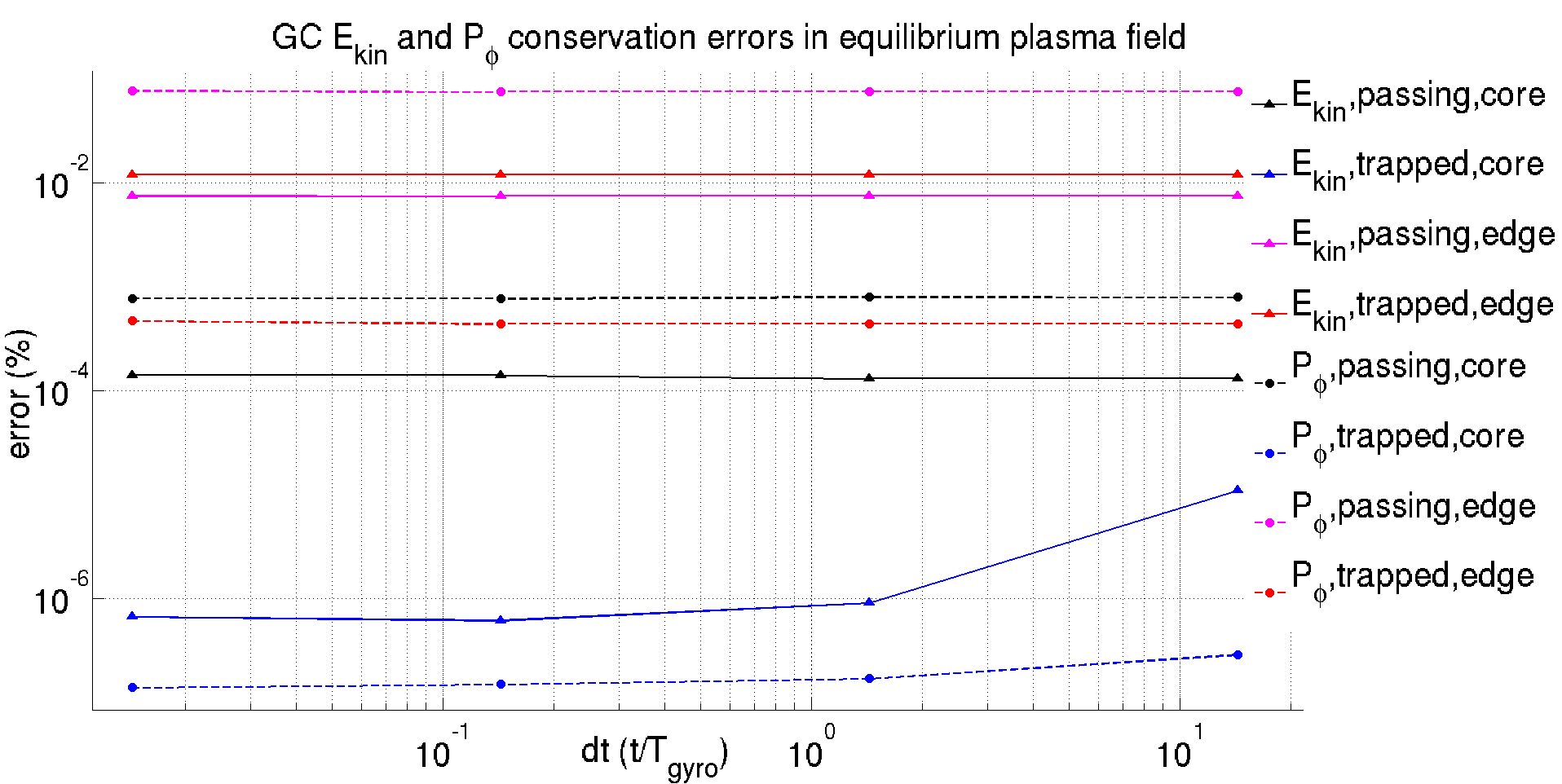}
\caption{GC $\mathrm{E_{kin}}$ and $\mathrm{P_{\phi}}$ error profiles for a 1ms simulation in stationary axisymmetric fields. The time step is normalised to the electron gyroperiod $\mathrm{T_{gyro}}$}
\label{fig:gc_Ekin_pphi_error_prof_eq}
\end{figure}

GC conservation errors for $\mathrm{E_{kin}}$ (plain lines) and  $\mathrm{P_{\phi}}$ (dashed lines) are shown in Figure \ref{fig:gc_Ekin_pphi_error_prof_eq} as a function of time step for the four test cases (core/edge, passing/trapped). The GC tracker performs clearly less well than the FO tracker, which is not surprising due to its lack of symplecticness and the presence of high order spatial derivatives in the equations of motion. In most cases, error convergence is reached for $\mathrm{\Delta t = 14 \cdot T_{gyro}}$. As for the FO tracker, conservation errors are much greater at the edge than in the core.

At this point, we turn our attention to the GC validity conditions. The quantities $\mathrm{\frac{\rho}{L_{{\nabla B}}}}$, $\mathrm{\frac{l_{\parallel}}{L_{{\parallel}_{\nabla B}}}}$ defined in Section \ref{guidingCenter} and the deviation of the magnetic moment $\mathrm{\mu}$ with respect to its mean value for the FO simulations (with $\mathrm{\Delta t = 1.4 \cdot 10^{-2} \cdot T_{gyro}}$) are reported in Table \ref{tab_eq_gc_validity_cond}. As expected (see Section \ref{guidingCenter}), electrons with energies of $\mathrm{10 MeV}$ in equilibrium fields satisfy the GC validity conditions. Note that the $\mathrm{\mu}$ variation given in Table \ref{tab_eq_gc_validity_cond} is essentially due to high frequency fluctuations around an approximately constant mean value, a behaviour which confirms the GC validity.


\begin{table}[b!]
\centering
\begin{tabular}{| c |  c |  c |  c |}
\hline
$\quad$ & $\mathrm{\frac{\rho}{L_{{\nabla B}}}}$  & $\mathrm{\frac{l_{\parallel}}{L_{{\parallel}_{\nabla B}}}}$ & $\mathrm{\frac{\max \left( \mu-<\mu> \right)}{\mu\left(t=0\right)}} \%$ \\
\hline
Core passing orbit & 5.1e-03 & 5.3e-02 & 7.8 \\
\hline
Core trapped orbit & 2.0e-03 & 2.9e-04 & 1.1e-01 \\
\hline
Edge passing orbit & 5.1e-03 & 7.0e-02 & 8.4 \\
\hline
Edge trapped orbit & 1.7e-03 & 2.7e-03 & 1.3e-01 \\
\hline
\end{tabular}
\caption{Estimation of critical quantities involved in GC validity conditions (see Section \ref{guidingCenter}) and magnetic moment variation for stationary axisymmetric test cases}
\label{tab_eq_gc_validity_cond}
\end{table}

\subsection{Tests in stationary non-axisymmetric fields} \label{magisland_verification}

\begin{figure}[h]
\centering
\includegraphics[width=10.0571cm, height=8cm]{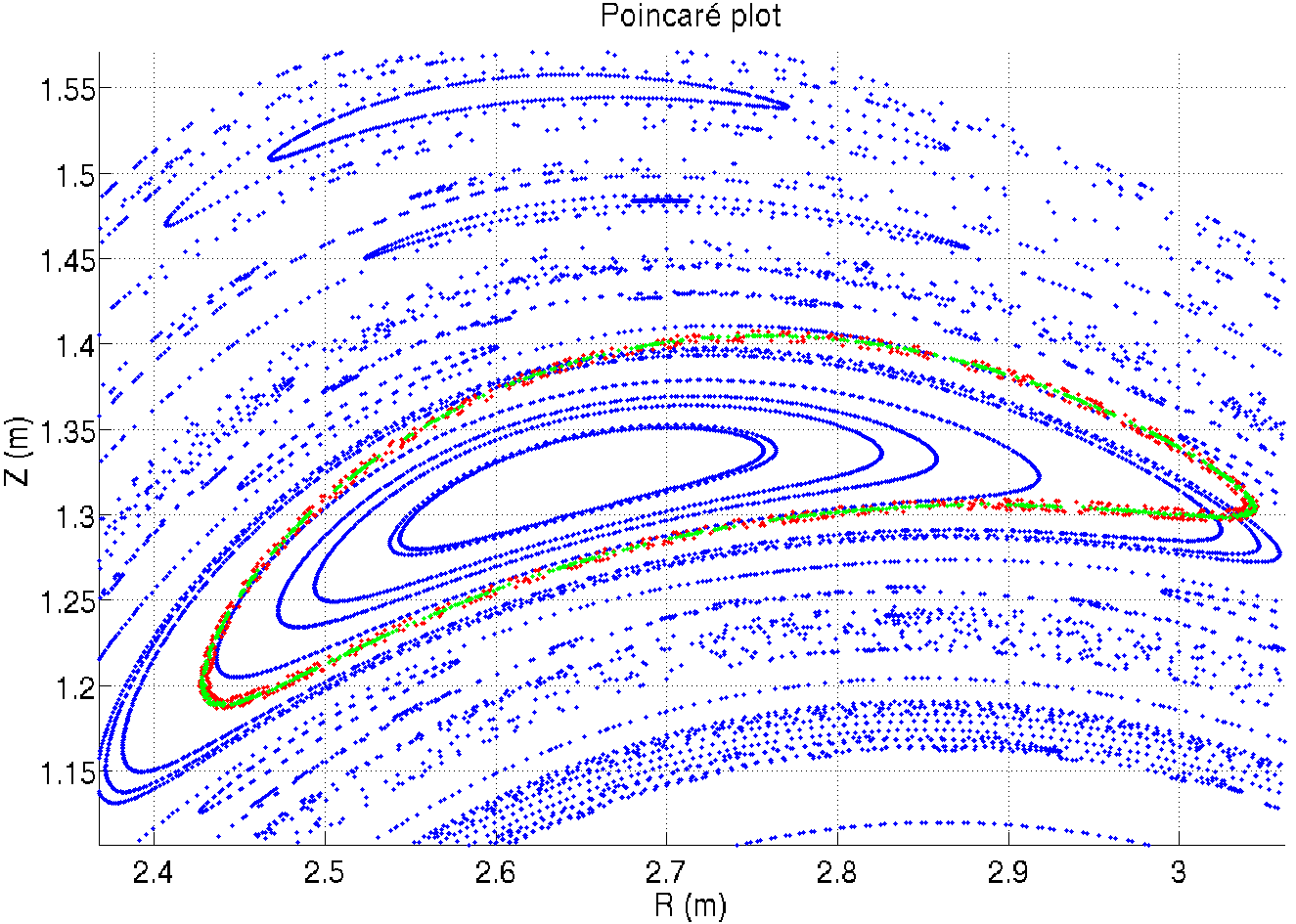}
\caption{Passing 10MeV orbit near q=2 surface in stationary non-axisymmetric fields. Red, green and blue dots denote respectively FO, GC and field line solutions}
\label{fig:tearing_passing_islandzoom}
\end{figure}

In the following, the JOREK fast particle tracker behaviour in non-axisymmetric stationary fields is described. Figure \ref{fig:tearing_passing_islandzoom} shows a Poincar\'e plot for a passing 10MeV electron near the q=2 surface. It can be seen that GC and FO solutions are consistent and that the orbit describes an m=2, n=1 island which is shifted from the magnetic island due to drifts (similarly to the axisymmetric case above), an effect already observed in \cite{papp111} and \cite{papp112}. 

\begin{figure}[h]
\centering
\includegraphics[width=10.5cm, height=6.5cm]{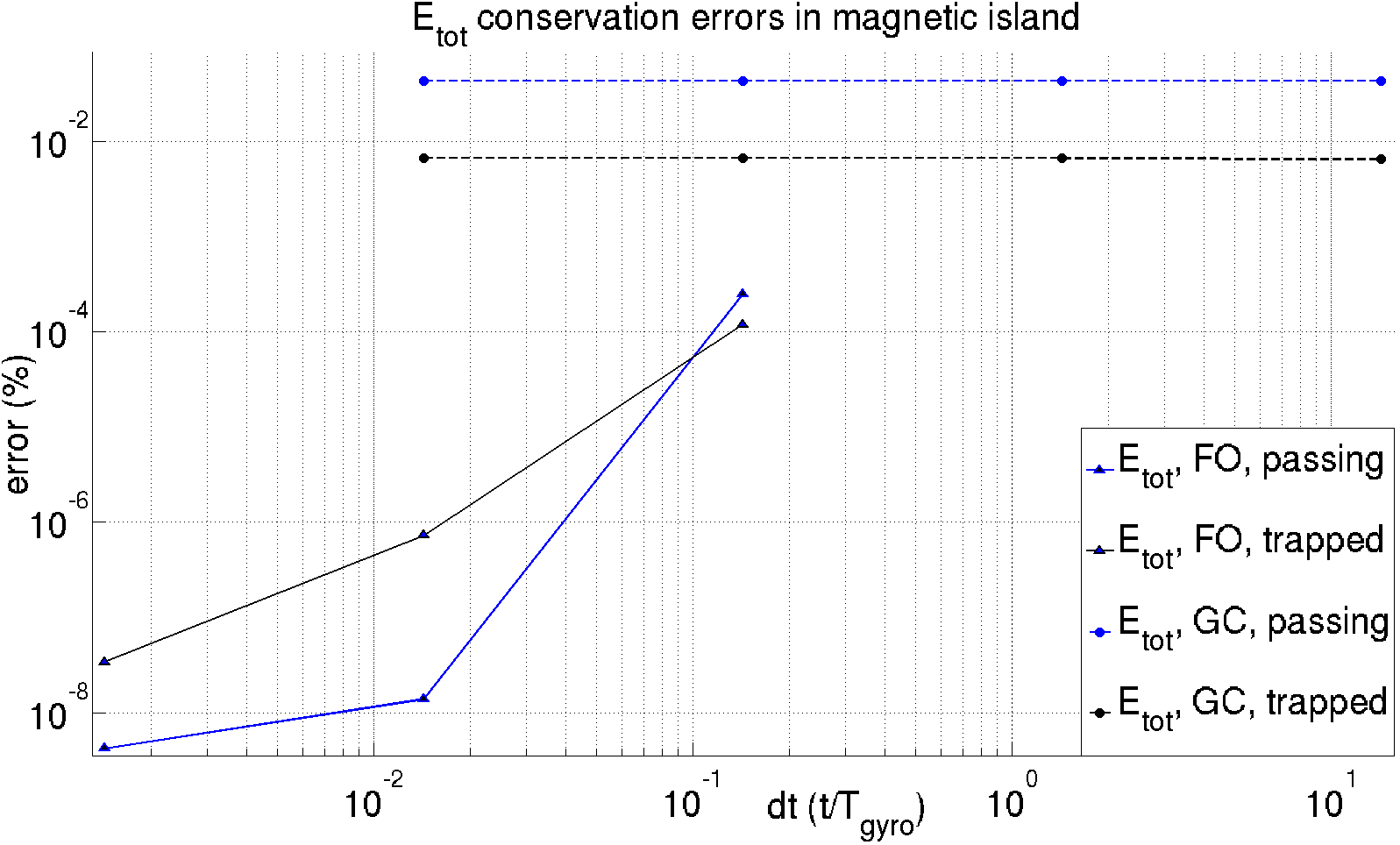}
\caption{GC and FO $\mathrm{E_{tot}}$ error profiles for a 1ms simulation in stationary non-axisymmetric fields. The time step is normalised to the electron gyroperiod $\mathrm{T_{gyro}}$}
\label{fig:fo_gc_Etot_error_prof_magisland}
\end{figure}

The FO and GC tracker maximum total energy $\mathrm{E_{tot}}$ conservation errors for a passing 10MeV and a trapped 1MeV electron near the q=2 surface in non-axisymmetric stationary fields are shown in Figure \ref{fig:fo_gc_Etot_error_prof_magisland}. As in the equilibrium case, the symplectic FO integrator shows much better conservation that the GC one at very small $\mathrm{\Delta t}$, but above $\mathrm{\Delta t = 1.4 \cdot 10^{-1} \cdot T_{gyro}}$, its performances degrade strongly. The GC integrator presents similar features to the above axisymmetric test case, although a general increase of the conservation errors can be observed, which is probably caused by the reduction of the fields smoothness compared to the axisymmetric case. Errors remain on the order of $10^{-2} \%$, which seems acceptable for the physics investigated below.


Critical quantities related to the GC validity conditions are shown in Table \ref{tab_magisland_gc_validity_cond} (calculated from the $\mathrm{\Delta t = 1.4 \cdot 10^{-2} \cdot T_{gyro}}$ FO simulation). The non-axisymmetry of the fields and the presence of islands does not cause a violation of the GC validity conditions. 


\begin{table}[h]
\centering
\begin{tabular}{| c |  c |  c |  c | c | c |}
\hline
  & $\mathrm{\frac{\rho}{L_{{\nabla B}}}}$ & $\mathrm{\frac{l_{\parallel}}{L_{{\parallel}_{\nabla B}}}}$  & $\mathrm{<\frac{E_{\parallel}}{E_{\perp}}>}$ & $\mathrm{\frac{\max \left(\mu-<\mu> \right)}{\mu \left(t=0 \right)}} \%$ \\
\hline
Passing particle & 5.2e-03 & 7.0e-02 & 4.4e-03 & 1.1e+01 \\
\hline
Trapped particle & 2.2e-03 & 1.7e-03 & 4.4e-03 & 1.5e-01 \\
\hline
\end{tabular}
\caption{Estimation of critical quantities involved in GC validity conditions (see Section \ref{guidingCenter}) and magnetic moment variation for stationary non-axisymmetric test cases}
\label{tab_magisland_gc_validity_cond}
\end{table}

Finally, a test of the total energy conservation in a stationary fully stochastic magnetic field is conducted for the GC and FO models using time steps of $\mathrm{\Delta t = 14 \cdot T_{gyro}}$ and $\mathrm{\Delta t = 0.014 \cdot T_{gyro}}$ respectively. This analysis is conducted tracking two electrons initialised in the plasma core region ($\mathrm{R=3.25m}$, $\mathrm{Z=0.22m}$, $\mathrm{\phi=45^{\circ}}$) and having kinetic energies of 1keV and 1MeV and pitch angles of (respectively) $\mathrm{\theta=170^{\circ}}$ (passing) and $\mathrm{\theta=100^{\circ}}$ (trapped), for a total simulation time of 1ms. The use of a kinetic energy of 1keV for the passing case, instead of 10MeV as in the above tests, is necessary in order to avoid a prompt deconfinement. The total energy conservation error is of $\mathrm{2.9\cdot 10^{-5}\%}$ (passing case) and $\mathrm{1.5 \cdot 10^{-2}\%}$ (trapped case) for the GC model and $\mathrm{1.8 \cdot 10^{-7}\%}$ (passing) and $\mathrm{2.1 \cdot 10^{-4}\%}$ (trapped) for the FO model, which is again acceptable for our purposes. It is worth mentioning that a small increment in the total energy is observed when a particle passes near the magnetic axis, which is a singular point of the JOREK mesh in this simulation, but this is small enough not to significantly affect the global energy conservation in these tests.

\section{Test electron transport in a JOREK-simulated MGI-triggered disruption in JET-like geometry} \label{physical_results}

This section presents the first physical application to RE physics of the above-described test particle module. The main objective is to assess to what extent a test electron population initialised before the TQ is deconfined by the MHD activity during a disruption, a question of high importance for the two mechanisms mentioned in the introduction: hot tail and Dreicer acceleration during the TQ. For this purpose, test electron populations will be initialised in the pre-TQ phase and tracked until the CQ phase. Initial energies going from 1keV to 10MeV will be considered in order to assess how transport depends on energy. Note that while 1keV electrons definitely exist before the TQ (the pre-TQ electron temperature in the simulated pulse is about 1keV), the 10MeV electrons in this transport study are purely hypothetical.

Here, since collisions are not yet implemented in the model, the objective is restricted to investigating collisionless transport properties. Since the mean free path of a 1keV electron (at a density of $10^{20}m^{-3}$) is of several hundreds of meter, which is much larger than $2 \pi q R \simeq 60$m in JET, the collisionless approximation appears sufficient to estimate the transport due to magnetic stochasticity. Note that, due to the lack of drag in the model, it has been chosen to cut the inductive $\mathrm{\left( \frac{\partial \psi}{\partial t}\right)}$ part of the electric field in the equations of motion in order to avoid a spurious acceleration of the electrons.

The JOREK simulation used here is a JET $\mathrm{D_2}$ MGI-triggered disruption simulation \cite{nardon17}\cite{fil2015} similar to the one used in Section \ref{numerical_tests} with the difference that here the q profile is artificially increased by $\sim$20\% (one consequence being the suppression of the internal kink mode). This particular simulation was chosen because it runs far enough into the CQ phase whereas the simulation from Section \ref{numerical_tests} suffers from numerical instabilities at the end of the TQ. It should be stressed that no RE were observed experimentally in this pulse. In fact, JOREK simulations of disruptions which produced a RE beam do not yet exist, although efforts have been started in this direction. But the present case is nonetheless interesting for two reasons: first, it is important to verify that the model does not predict RE when they are not observed; second, qualitative findings made on the present case may have a more general importance. It is clear nevertheless that the present study is only a first step and that many more simulations will be necessary if we are to gain a deep understanding of RE physics during disruptions using JOREK.

\begin{figure}[h]
	\centering
	\subfigure[A][$\mathrm{0<t(ms)<0.1}$] {\includegraphics[width=4.0cm, height=7cm]{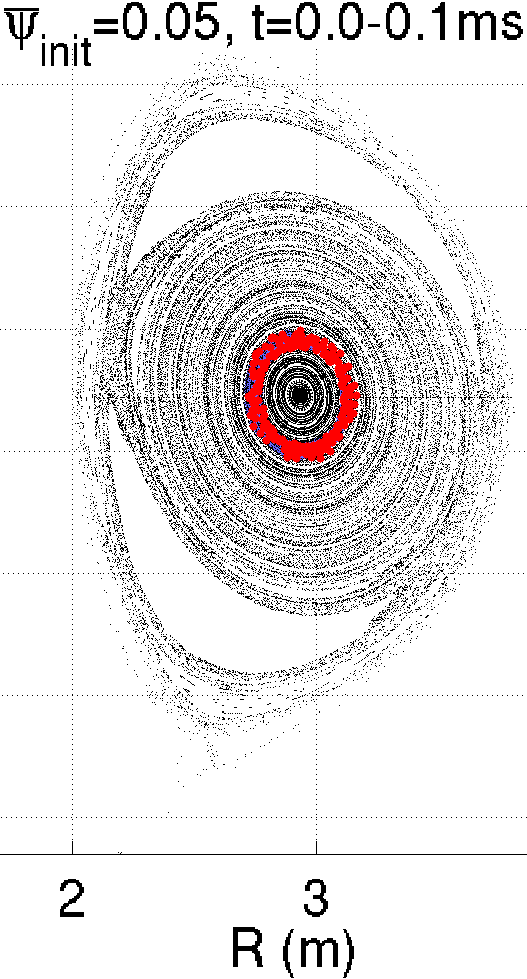} \label{fig:pseudo_psi0e05_1}}
	\subfigure[B][$\mathrm{0.405<t( ms)<0.505}$] {\includegraphics[width=4.9cm, height=7cm]{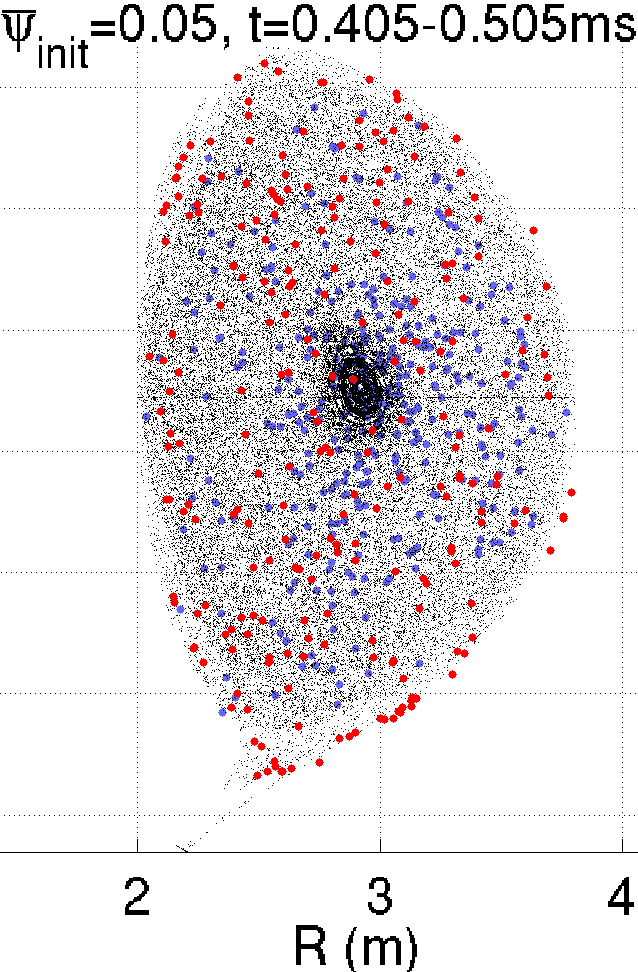} \label{fig:pseudo_psi0e05_2}}
	\subfigure[C][$\mathrm{3.255<t(ms)<3.355}$] {\includegraphics[width=4.75cm, height=7cm]{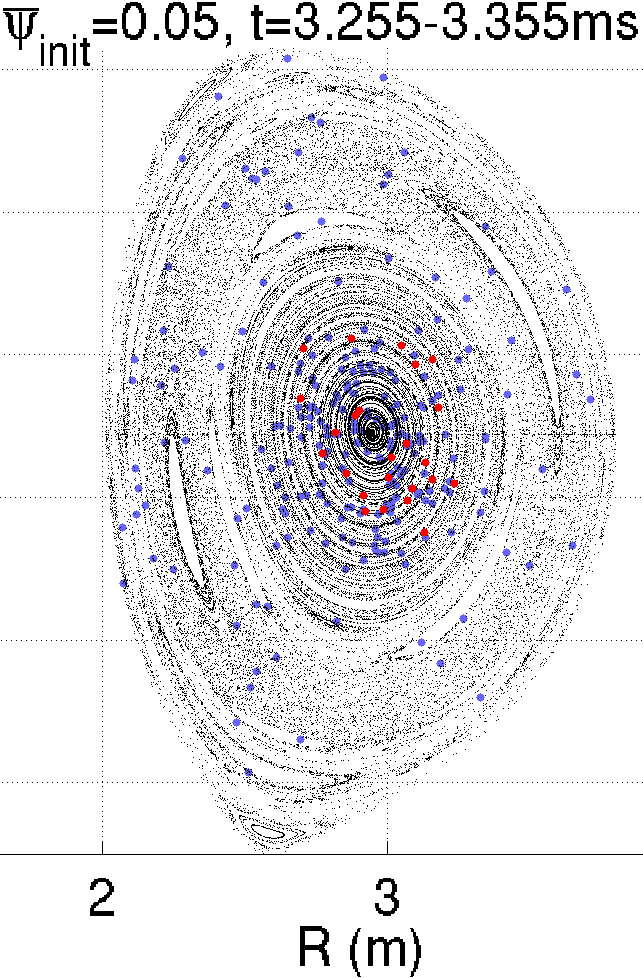} \label{fig:pseudo_psi0e05_3}}
	\caption{Pseudo-Poincar\'e plots at different times in the simulation. Blue (resp. red) dots are electrons with an initial kinetic energy of 1keV (resp. 10MeV), while black dots are field lines.}
	\label{fig:pseudo_poincare_psi0e05_psi0e07}
\end{figure}

Let us start by briefly describing the evolution of the magnetic field structure in this disruption simulation. Figure \ref{fig:pseudo_poincare_psi0e05_psi0e07} presents Poincar\'e plots of the magnetic field lines (FL) (black dots) at 3 times in the simulation (note that in this section, t=0 corresponds to a time during the pre-TQ phase, about 0.5ms before the TQ). It can be seen that the pre-TQ phase (left) is dominated by the growth of a large $\mathrm{m=2}$, $\mathrm{n=1}$ magnetic island, while the TQ (middle) is characterized by a global stochastisation of the magnetic field after the stochastic region has expanded from the edge to the core of the plasma. Finally, during the CQ (right), closed flux surfaces gradually reappear, firstly in the core and subsequently at the edge. We call attention on the fact that this evolution is different from the one assumed by Boozer and Punjabi in \cite{boozer16}, where a broad stochastic region is initially bounded by an annulus of closed flux surfaces which are progressively destroyed in time. This does not imply that Boozer and Punjabi made a wrong assumption, because different scenarios may be possible. In fact, NIMROD simulations \cite{izzo11}\cite{izzo12} display various types of evolutions depending on, for example, machine size, divertor vs. limiter configuration, existence and structure of the 1/1 mode, etc. The scenario studied in the present paper should by no means be considered universal. 

\begin{figure}[h]
\centering
	\subfigure{\includegraphics[width=12.25cm, height=5.5cm]{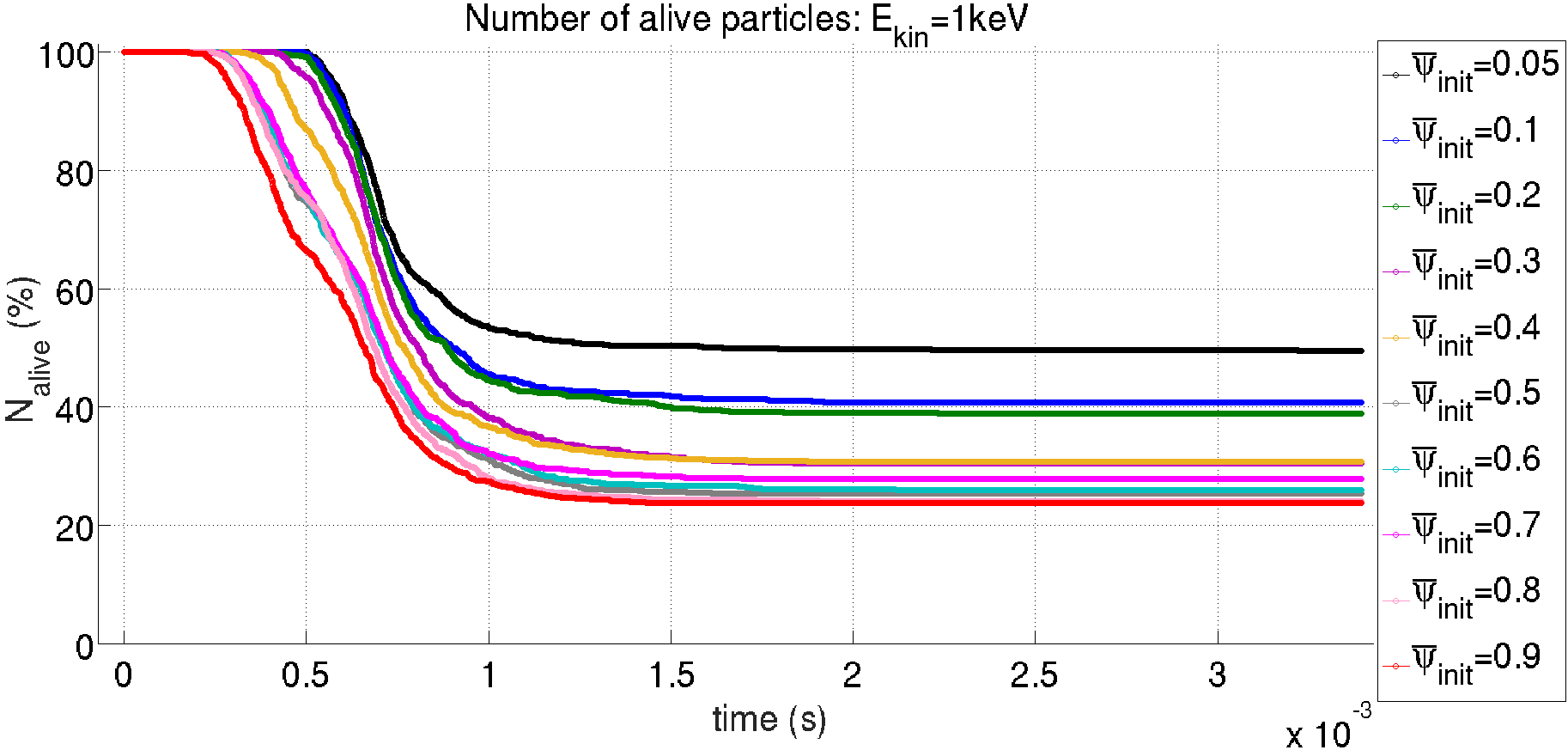} \label{fig:loss_1kev_time}}
	\subfigure{\includegraphics[width=12.25cm, height=5.5cm]{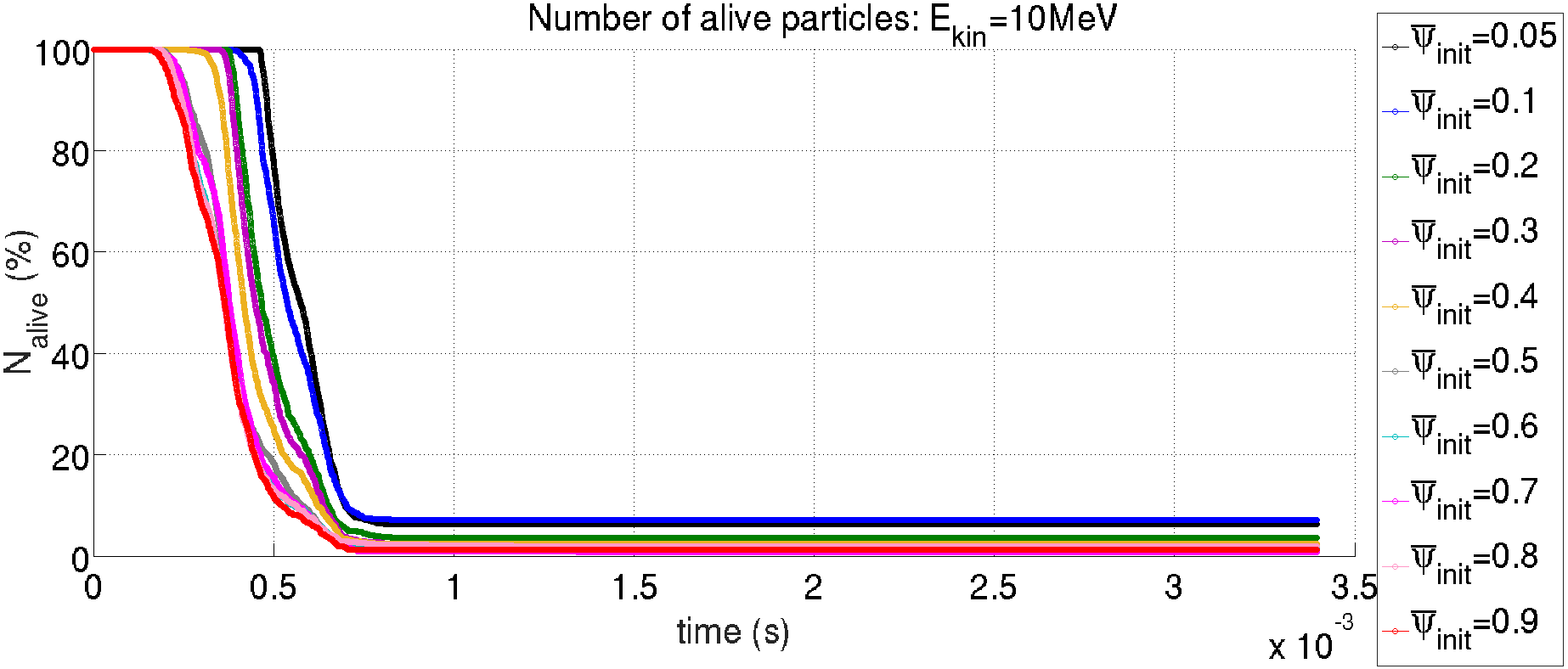} \label{fig:loss_10Mev_time}}
\caption{Fraction of electron (GC model) populations which are still in the plasma at a given time. Each population contains 1000 electrons initialised at 1keV (above) or 10MeV (below) and at a given radial position (different colors show different initial radii).}
\label{fig:time_loss_profiles}
\end{figure}

Figures \ref{fig:pseudo_poincare_psi0e05_psi0e07} and \ref{fig:time_loss_profiles} present the results of simulations where populations of 1000 test electrons are initialised near given flux surfaces $\mathrm{\bar{\psi}}=\mathrm{\bar{\psi}_{init}}$ (where $\mathrm{\bar{\psi}}$ is the n=0 component of the poloidal flux normalised to be 0 at the magnetic axis and 1 at the plasma edge) in the pre-TQ phase and tracked with the GC model for $\approx$ 3.4ms (time step $\mathrm{=14 \cdot T_{gyro}}$), i.e. until the CQ phase. The initial kinetic energy of the electrons is 1keV (blue dots in Figure \ref{fig:pseudo_poincare_psi0e05_psi0e07} and upper plot in Figure \ref{fig:time_loss_profiles}) and 10MeV (red dots in Figure \ref{fig:pseudo_poincare_psi0e05_psi0e07} and lower plot in Figure \ref{fig:time_loss_profiles}). The initial pitch angle is $\mathrm{\theta =170^{\circ}}$ (passing electrons) which is chosen within the typical experimental interval, seen in various machines, of $\mathrm{\approx[5,12]^{\circ}}$ \cite{jaspers01}\cite{zychen06}\cite{stahl13} and respecting the runaway electron counter-current motion \cite{izzo11}\cite{papp111} (in JET, plasma current and magnetic field are both in clockwise direction seen from above so a counter-B field momentum is required for a correct RE initialisation). We refer to Figure \ref{fig:pseudo_poincare_psi0e05_psi0e07} as ``pseudo-Poincar\'e'' plots. These plots are obtained by representing the nearest position, from the discrete trajectory output by the code, of each electron to a given poloidal plane within a short time $\mathrm{\left( \delta t = 0.1ms \right)}$ and toroidal angle $\mathrm{\left( \delta \phi= \pm 30^{\circ}\right)}$ window. Note that for magnetic FL, these plots are standard Poincar\'e plots.

Figure \ref{fig:time_loss_profiles} shows the fraction of electrons which are still ``alive'', i.e. which have not been lost, after a given time, for initial energies of 1keV (upper plot) and 10MeV (lower plot) and a set of initial radii (different colors). It can be seen that electrons (both at 1keV and 10MeV) start being lost at a time which varies between 0.25ms, for electrons initialised at the edge, and 0.5ms, for those initialised in the core. This corresponds to the gradual stochastisation of the magnetic field during the TQ, which starts from the edge near 0.25ms and reaches the core at about 0.5ms (as visible in the middle plot in Figure \ref{fig:pseudo_poincare_psi0e05_psi0e07}). Electron losses are faster for 10MeV than 1keV electrons. However, around 0.75ms, losses stop for the 10MeV population. This corresponds to the reappearance of flux surfaces in the core of the plasma, which trap the electrons located in this region. It can indeed be seen in the right plot of Figure \ref{fig:pseudo_poincare_psi0e05_psi0e07} that the electrons from the 10MeV population which are still present at the end of the simulation are located in the core region. In contrast, 1keV electrons are present throughout the plasma at the end of the simulation. The difference is related to the fact that 10MeV electrons diffuse across the stochastic region much faster than 1keV ones due to their faster parallel motion (as will be shown below), and that the latter diffuse slowly enough that a significant part of them remains in the stochastic region until a flux surface reappears at the edge and stops the loss process. This explains why 1keV electron losses stop at a later time than 10MeV losses (see Figure \ref{fig:time_loss_profiles}). The final fraction of remaining electrons is much larger for 1keV (a few tens of \%) than 10MeV, although the fraction of 10MeV ``survivors'' is not negligible: typically a few \%. 

\begin{figure}[h]
	\centering
	{\includegraphics[width=12.25cm, height=5.5cm]{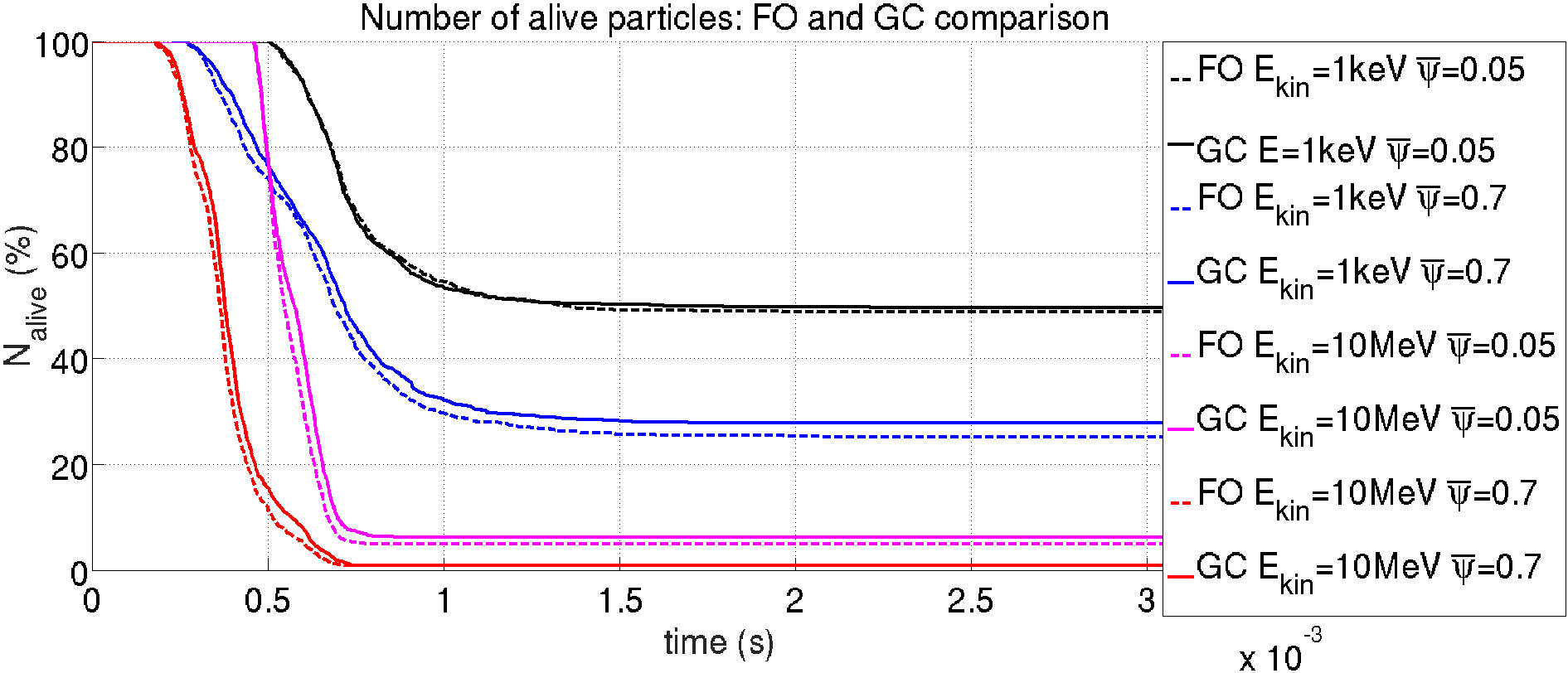} \label{fig:loss_FO_GC_comp}}  
	\caption{Comparison among evolutions of ``surviving'' electron populations obtained using FO (dashed lines) and GC (solid lines) trackers (different colors are related to different test cases).}
	\label{fig:FO_vs_GC_loss_profiles}
\end{figure}

Figure \ref{fig:FO_vs_GC_loss_profiles} shows, for a few selected cases, that the GC and FO models agree rather well in terms of electron losses (for FO simulations, the gyro-angle was initialized randomly in the interval $\mathrm{\left[0,2\pi \right)}$ and a time step of $\mathrm{0.014 \cdot T_{gyro}}$ was used). This gives confidence that the computationally much faster GC model can be used for the type of studies presented in this paper. It is important to note however that for other aspects of RE physics, such as synchrotron radiation or the analysis of electron orbits for $\mathrm{E_{kin} > 10MeV}$, FO effects may have to be taken into account, as reported in \cite{wang16} and \cite{carbajal17}.     

\begin{figure}[h]
\centering
\includegraphics[width=8.5cm, height=4.75cm]{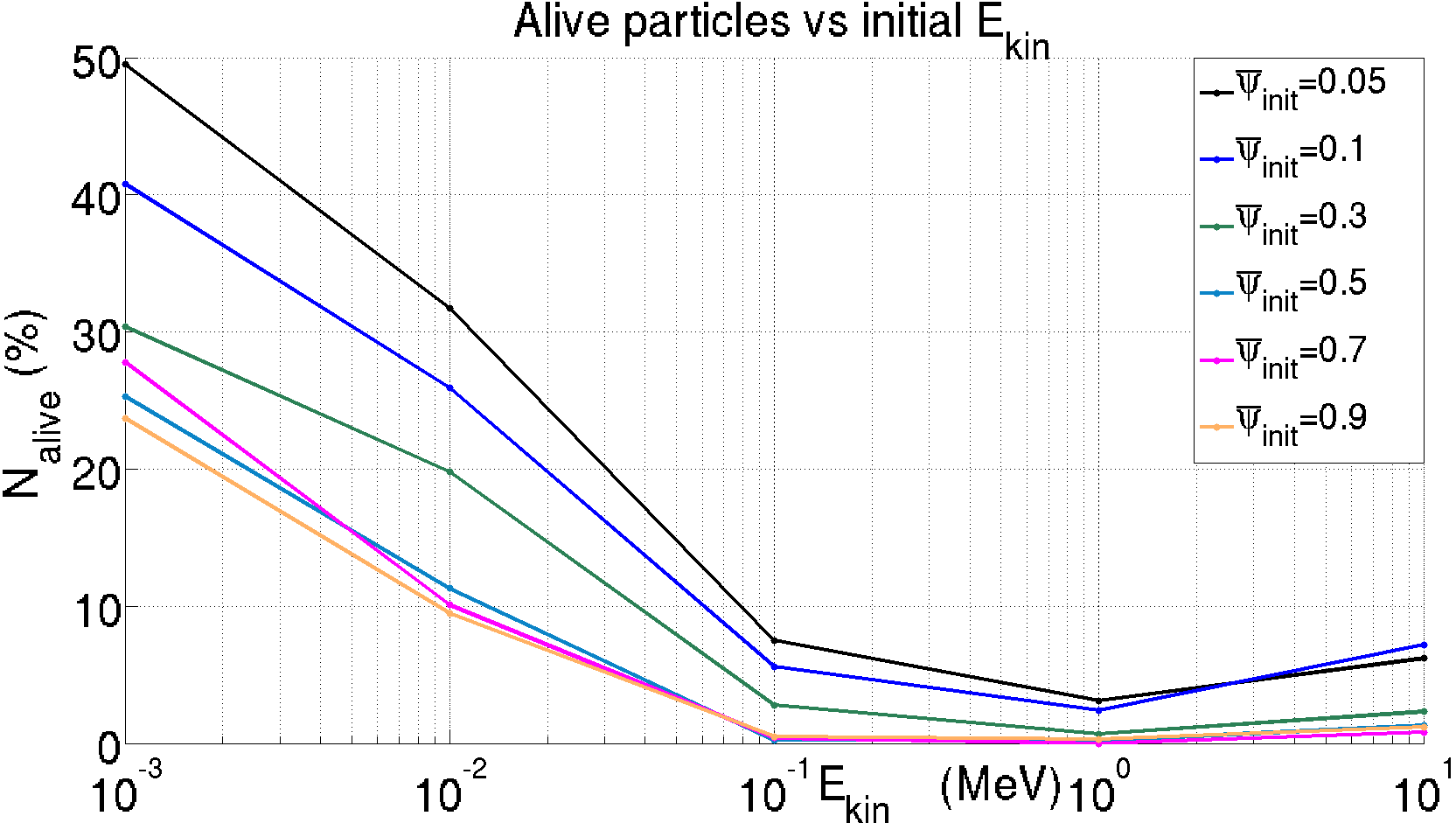} 
\label{fig:loss_vs_enegy_init}
\caption{Fraction of electrons (GC model) still confined at the end of the simulation versus initial kinetic energy and initial radial position (different colors)}
\label{fig:loss_vs_initial_condition}
\end{figure}

Figure \ref{fig:loss_vs_initial_condition} shows how the fraction of remaining electrons (or ``survivors'') at the end of the simulation depends on the initial energy and radial position. It can be seen that the fraction of survivors decreases when the initial energy increases from 1keV to 1MeV with a saturation-like behaviour above 100keV, and then, interestingly, increases above 1MeV. The behaviour below 1MeV is qualitatively consistent with the picture that electrons are lost by parallel transport along stochastic FL, because increasing energy means increasing parallel velocity. The saturation above 100keV is likely related to the fact that the parallel velocity tends to the speed of light. Finally, the fact that losses decrease with energy above 1MeV is consistent with the phase-averaging effect related to the orbit shift at high energy already mentioned in Section \ref{numerical_tests}, as described by \cite{mynick81} and \cite{abdullaev16}.



\begin{figure}[h]
\centering
	{\includegraphics[width=10.0cm, height=5.5cm]{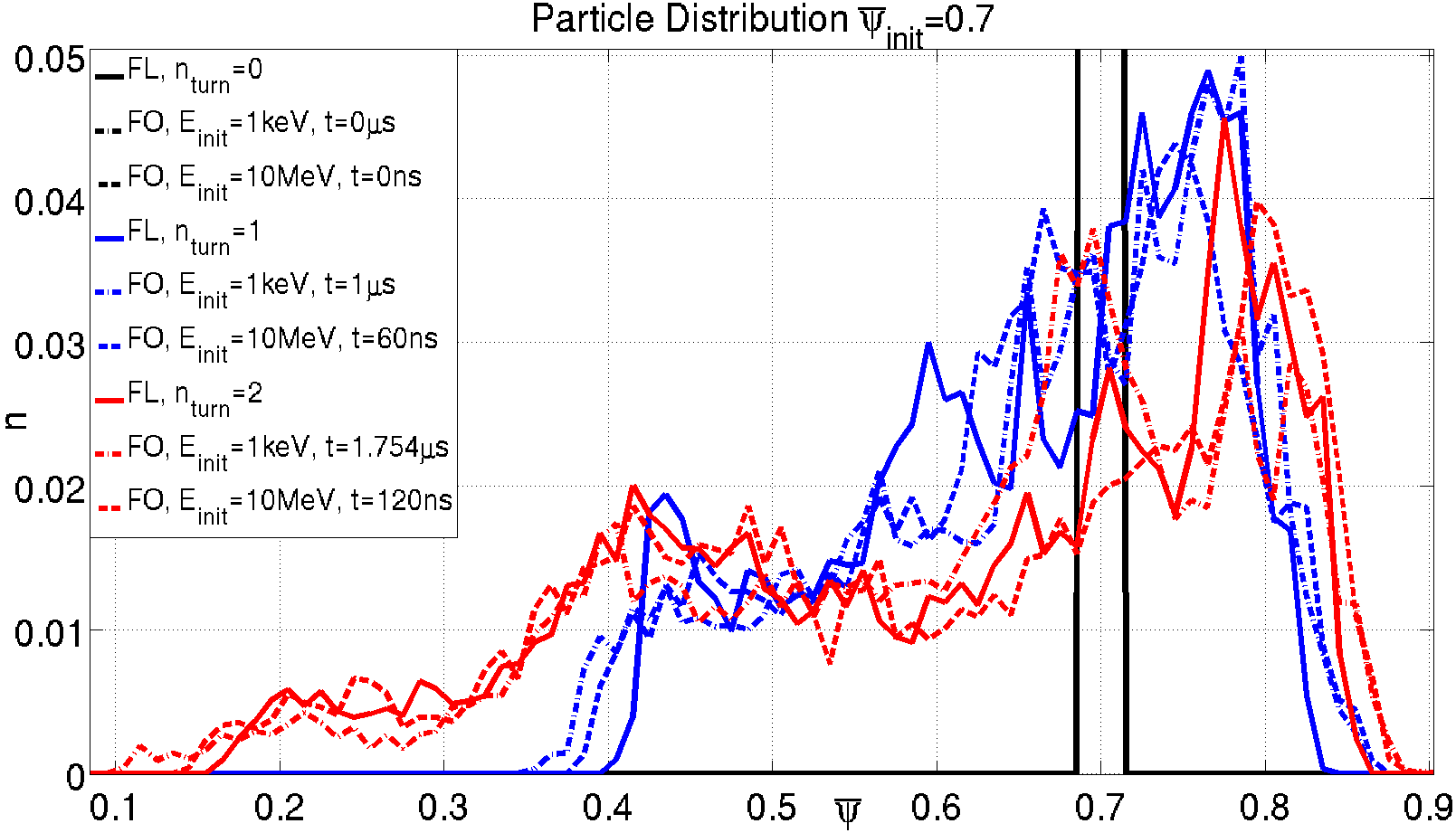} \label{fig:fl_FO_distributions_psi0e7}}
	{\includegraphics[width=10.0cm, height=5.5cm]{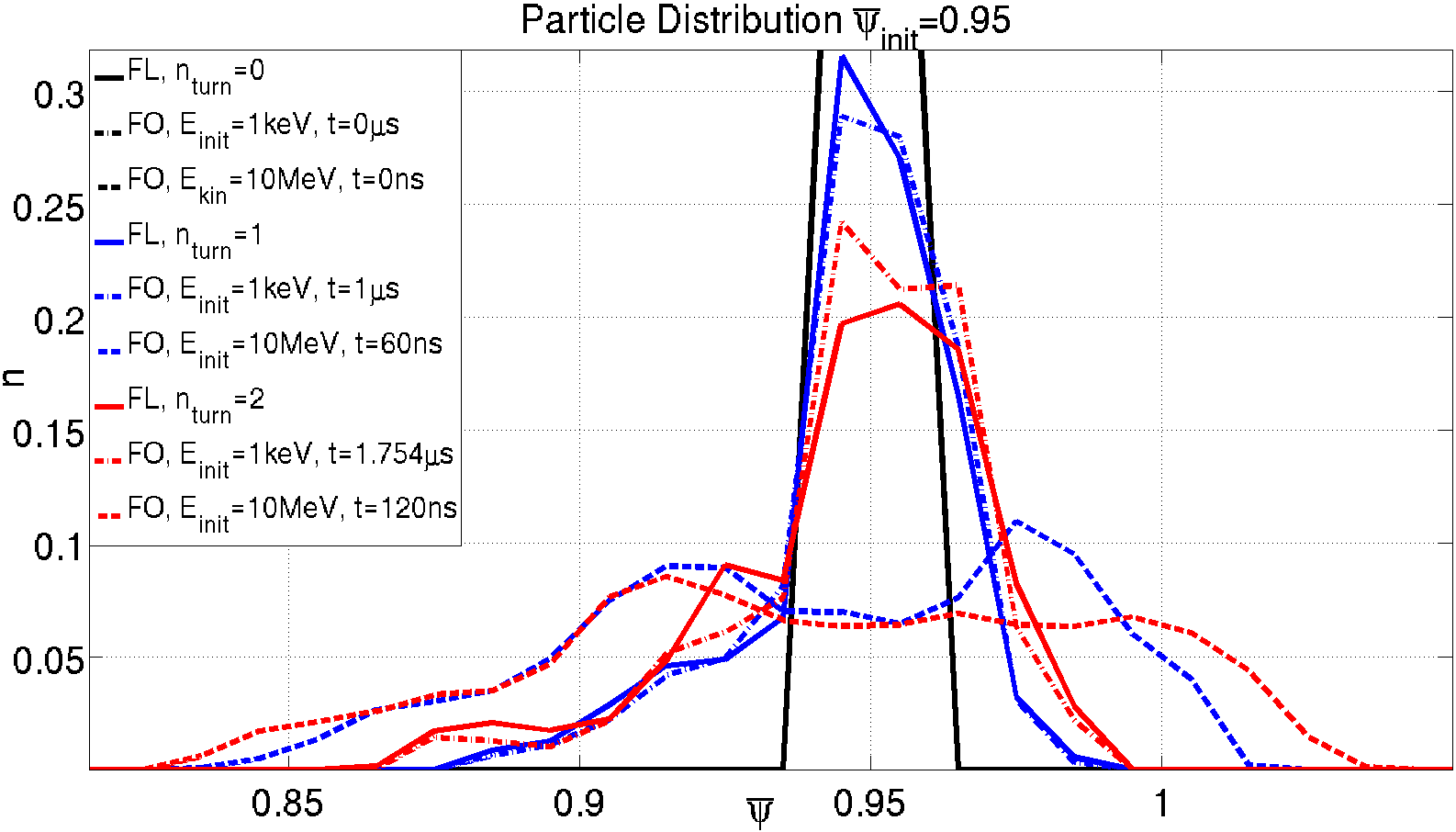} \label{fig:fl_FO_distributions_psi0e95}}
\caption{Distributions of field lines (solid lines) and electrons initialised at 1keV (dash-dotted lines) and 10MeV (dashed lines) after 0, 1 and 2 toroidal turns, starting from narrow Heaviside-like distributions centered at $\mathrm{\bar{\psi}_{init}=0.7}$ (upper plot) and $\mathrm{\bar{\psi}_{init}=0.95}$ (lower plot).}
\label{fig:fl_FO_distributions}
\end{figure}

\begin{figure}[h]
\centering
	{\includegraphics[width=5.5cm, height=6.75cm]{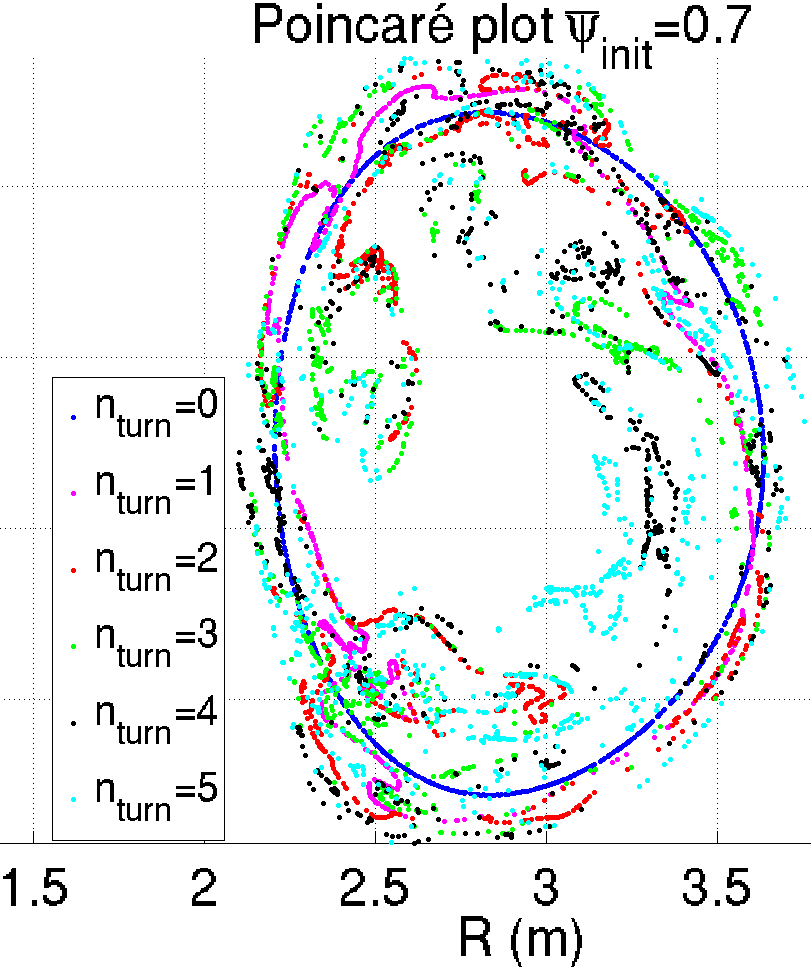} \label{fig:fl_poincare_psi0e7_1}}
	{\includegraphics[width=5.0cm, height=6.75cm]{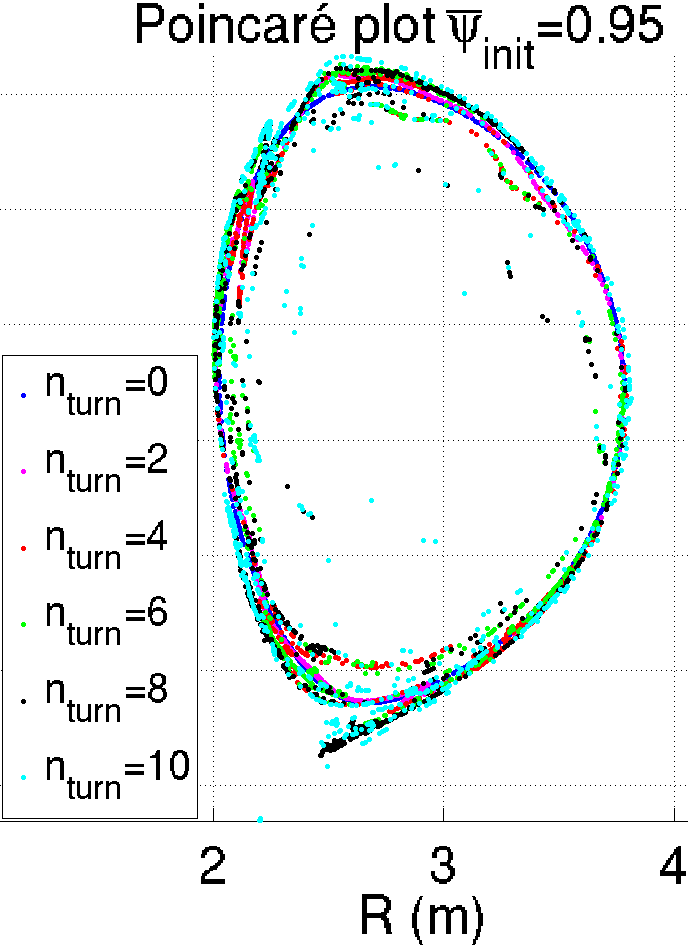} \label{fig:fl_poincare_psi0e95_1}}
\caption{Poincar\'e plots (at a toroidal angle of $45^{\circ}$) showing evolutions of field line populations initialised at $\mathrm{\bar{\psi}_{init}=0.7}$ (left plot) and $\mathrm{\bar{\psi}_{init}=0.95}$ (right plot). Different colors represent field line positions after an increase of one (left plot) or two (right plot) toroidal turns}
\label{fig:fl_poincare_evolution_psi0e7_psi0e95}
\end{figure}

Figure \ref{fig:fl_FO_distributions} gives information on local (in time and space) transport properties. It shows how initial narrow Heaviside-like radial distributions of FL or electrons (width $\mathrm{\Delta \bar{\psi}_{init}=10^{-3}}$) evolve after a given number of toroidal turns (electron distributions are represented at times corresponding to a given number of toroidal turns). Populations are initialised at t=0.47ms (i.e. during the TQ, when the magnetic field is stochastic across the whole plasma - see middle plot in Figure \ref{fig:pseudo_poincare_psi0e05_psi0e07}) and at two radial positions: $\mathrm{\bar{\psi}_{init}=0.7}$ (upper plot) and $\mathrm{\bar{\psi}_{init}=0.95}$ (lower plot), and two types of electron populations are tracked: one initially at 1keV and the other at 10MeV. Different colors represent different numbers of turns between 0 and 2. Each population is made of $10^4$ FL or electrons. A first observation on Figure \ref{fig:fl_FO_distributions} is that distributions of FL and electrons initialised at $\mathrm{\bar{\psi}_{init}=0.7}$ are very similar after a given number of toroidal turns. This strongly suggests that electron radial diffusion is essentially due to parallel motion along the stochastic FL with a magnitude proportional to the parallel velocity, as discussed in \cite{isichenko911} \cite{isichenko912}. The situation is less clear for FL and electrons initialised at $\mathrm{\bar{\psi}_{init}=0.95}$. Indeed, while FL and 1keV electrons have comparable distributions after a given number of turns, 10MeV electron distributions show clear differences. These are likely due to orbit shift effects. A second observation on Figure \ref{fig:fl_FO_distributions} is that distributions initialised at $\mathrm{\bar{\psi}_{init}=0.7}$ spread radially much faster than those initialised at $\mathrm{\bar{\psi}_{init}=0.95}$ (which probably explains why orbit shift effects are visible only in the latter case). Radial transport is therefore clearly not homogeneous within the plasma. This can be seen also in the Poincar\'e plots shown in Figure \ref{fig:fl_poincare_evolution_psi0e7_psi0e95}. In these plots, a large number of FL are initialised at $\mathrm{\bar{\psi}_{init}=0.7}$ (left) and $\mathrm{\bar{\psi}_{init}=0.95}$ (right) and different colors represent FL positions after different numbers of toroidal turns. The radial spread is obviously much faster for $\mathrm{\bar{\psi}_{init}=0.7}$ than $\mathrm{\bar{\psi}_{init}=0.95}$, consistently with Figure \ref{fig:fl_FO_distributions}. It is likely that the relatively slow transport at the edge plays an important role in global electron confinement properties. A side remark on Figure \ref{fig:fl_poincare_evolution_psi0e7_psi0e95} is that transport in the core does not seem to result from a quasi-linear diffusion where information on the phase (the initial poloidal or toroidal angle) is lost in a time much shorter than the radial diffusion time.

\section{Discussion and future plans} \label{conclusions}

In this work, the JOREK module capable of computing relativistic full and GC orbits in time-varying 3D MHD fields is presented. A volume preserving symplectic scheme is used for FO tracking while the Cash-Karp Runge-Kutta method with time step control is used for GC tracking. The module was verified in both stationary axisymmetric and non-axisymmetric fields, showing good invariants of motion conservation properties up to a physical time of 1ms. 

The JOREK fast particle tracker was used for studying electron confinement in an MGI-triggered disruption simulation in JET-like geometry \cite{nardon17}\cite{fil2015}. Results suggest that electron deconfinement due to magnetic stochasticity does not prevent the hot tail mechanism or the Dreicer mechanism during the TQ. Indeed, a fraction of the order of 1\% up to a few tens of \% of a test electron population initialised before the TQ is typically not deconfined, depending on the initial energy and position. This fraction should be regarded as a large number since very small RE densities are sufficient to carry the whole plasma current (this is all the more true in presence of a strong amplification by the avalanche mechanism during the CQ). On the other hand, as mentioned in Section \ref{physical_results}, the simulated pulse did not produce RE, which appears paradoxical.

A possible reason could be that the MHD activity simulated by JOREK is too weak and therefore that electron losses are underestimated in the present work. In fact, it has been stressed in \cite{nardon17} that the plasma current spike in the simulations is typically one order of magnitude smaller than in the experiment, which points in the same direction. The edge region deserves particular attention due to its relatively low stochastic transport (see Section \ref{physical_results}). It is planned to investigate (by means of JOREK-STARWALL simulations with a resistive wall model \cite{hoelzl12}\cite{merkel15}) whether this is physical or whether the fixed $\psi$ boundary condition used here, with a computational boundary closer to the plasma than the actual wall, artificially reduces magnetic stochasticity at the edge. More generally, future efforts will focus on validating JOREK simulations versus magnetic measurements. Until this is achieved, conclusions on fast particle confinement should be taken with caution.

A quantitative validation of the test particle module itself appears to be a difficult task which may require the implementation of synthetic diagnostics (soft X-ray for example). Qualitative validation seems a more realistic goal in the near term. For example, it is planned to assess whether the model is qualitatively consistent with the RE existence domain observed at JET versus toroidal field and Argon/$\mathrm{D_2}$ fraction in the injected gas during MGI experiments \cite{reux15}. The observed trend that RE appear more easily in limiter than divertor configuration \cite{reux15} can also provide a good test. Moreover, the recently developed RE beam tomographic reconstruction \cite{ficker17}, might be used for checking the agreement between the spatial distribution of surviving electrons in simulations and the observed beam geometry. 

An important ongoing development of the test particle module will be the introduction of collisional drag terms, which will allow direct investigations of primary RE generation mechanisms.

Finally, it is also planned to perform test particle studies in different types of disruptions simulated by JOREK, e.g. disruptions triggered by shattered pellet injection, where first JOREK simulations show clear differences in terms of MHD activity compared to MGI-triggered disruption simulations \cite{hu2017}.

\section{Acknowledgments} \label{Acknowledgments}

The authors wish to express their gratitude to Konsta Sarkimaki and Taina Kurki-Suonio for their collaboration on testing the JOREK fast particle tracker aginst ASCOT code results in equilibrium fields. Moreover, we wish to thank Xavier Garbet, Philippe Ghendrih, Gergely Papp, Eero Hirvijoki, David Pfefferl\'e, Alain J. Brizard and Allen H. Boozer for their advice and fruitful discussions.

This work has been carried out within the framework of the EUROfusion Consortium and has received funding from the Euroatom research and training programme 2014-2018 under grant agreement No 633053. The views and opinions expressed herein do not necessarly reflect those of the European Commission.

\bibliographystyle{ieeetr}
\bibliography{bibliography}




\end{document}